\documentclass[11pt]{article}
\usepackage{latexsym}

\begin{document}

\begin{titlepage}
\vspace{.7in}
\begin{center}
\Large\bf {SCALE-INVARIANT GRAVITY: PARTICLE DYNAMICS}\normalfont

\vspace{.7in}

\large{Julian Barbour}

\vspace{.4in}

\normalsize{College Farm, South Newington, Banbury, Oxon, OX15
4JG, UK}

\vspace{.4in} Electronic address: julian@platonia.com

\end{center}

\begin{abstract}

A new and universal method for implementing scale invariance, called \textit{best matching}, is presented.
It extends to scaling the method introduced by Bertotti and the author to create a fully relational dynamics
that satisfies Mach's principle. The method is illustrated here in the context of non-relativistic
gravitational particle dynamics. It leads to far stronger predictions than general Newtonian dynamics. The
energy and angular momentum of an `island universe' must be exactly zero and its size, measured by its moment of inertia, cannot change. This constancy is enforced because the scale invariance requires all potentials to be
homogeneous of degree -2. It is remarkable that one can nevertheless exactly recover the standard observed
Newtonian laws and forces, which are merely accompanied by an extremely weak universal force like the one due to Einstein's
cosmological constant. In contrast to Newtonian and Einsteinian dynamics, both the gravitational constant G and 
the strength of the cosmological force are uniquely determined by the matter distribution of the universe. Estimates
of their values in agreement with observations are obtained. Best matching implements a \textit{dynamics of pure 
shape} for which the action is a dimensionless number. If the universe obeys such scale-invariant law, steadily increasing inhomogeneity, not expansion of the universe, causes the Hubble red shift. The application of best
matching to geometrodynamics is treated in a companion paper.

\end{abstract}

\end{titlepage}

\section{Introduction}

The Euclidean symmetries of three-dimensional space
(\textit{3-space}) do not fully determine Newtonian kinematics.
Absolute time and inertial frames are puzzling extra
structures. They are eliminated in this paper in a dynamics that
is \textit{fully invariant} under Euclidean translations,
rotations \textit{and} dilatations. Besides results of independent
interest, this paper introduces the techniques of the companion
paper \cite{ABFO}, which modifies Einstein's general relativity (GR), making it
scale invariant in a similar manner. Both papers extend the
\textit{3-space approach} of \cite{BOF, AB}, which derives
spacetime kinematics, including the universal light cone and the
gauge principle, from 3-space first principles.

Weyl \cite{Weyl1} made the first serious attempt to create a
scale-invariant (spacetime) dynamics, but his use of a conformally
transformed 4-vector field to compensate conformal transformations
of the gravitational field came to grief. Einstein \cite{Einstein}
noted that it would make atoms emit spectral lines with
history-dependent frequencies. This contradicted the observed
facts. Many years later, Dirac \cite{DiracSI} revived Weyl's 
idea in a simplified form in which the compensating 4-vector field 
is replaced by a simpler scalar
compensating field. In his theory, there is no longer the flagrant disagreement
with observation associated with Einstein's spectral argument. Nevertheless,
Einstein would probably still have had reservations, since in Dirac's theory
(which is a form of Brans--Dicke theory \cite{BD}) the coupling constants of the 
various forces of nature are epoch and location dependent. The strong equivalence
principle (SEP), according to which the laws of nature, including the values
of coupling constants, must be exactly the same everywhere and at every epoch,
is violated. Dirac regarded this as a virtue,
since he hoped it would justify his Large Numbers Hypothesis, but others would
prefer the rigour of the SEP. From the late eighties, numerous authors
have proposed Dirac-type theories \cite{Wetterich}. The scalar compensating field
is now widely called the dilaton, and it has been conjectured \cite{Wetterich} that
its vacuum expectation value determines the Planck length in an emergent fashion.
A dilaton field is a necessary concomitant of the Einsteinian graviton in string theory,
and there is now a vast literature on related matters.

There are reasons to be unhappy about the present situation. The general principles
of physics impose few restrictions on scalar fields, so the scope for `ad hocery'
in order to match the latest observations is great. Even in string theory, the many different compactification
possibilities lead to different dilatonic
properties, and there seems to be little hope at this stage of unambiguous predictions for the form
of the resulting effective four-dimensional theory. In addition, quantization, which 
often fails to respect classical scale invariance, introduces further uncertainty
in the form of anomalies and arbitrary cutoffs.  

In this and the companion paper \cite{ABFO}, a new and universal approach to scale invariance is proposed. 
It makes no use of compensating fields at all and leads directly and naturally
to a mechanism whereby
the strong equivalence principle can be maintained. Since scalar fields play no essential role
in the theory, there is much less scope for fudging. Scale-invariant gravity makes at least some unambiguous predictions, and it is therefore decidedly vulnerable to experimental disproof.

The idea behind scale-invariant gravity is in fact closely related to Weyl's original
inspiration, which was to implement Descartes's dream of explaining 
all of physics by geometry. In 1919, despite Einstein's reservations, 
Weyl wrote \cite{Weyl2}: ``Descartes's dream
 of a purely geometrical physics is fulfilled in a remarkable way, though admittedly
 not one that he could have foreseen. In its concepts, physics in no way goes 
 beyond geometry.'' However, years later he admitted defeat and wrote \cite{Weyl3}
  ``the facts of atomism teach us that
length is not relative but absolute {\dots} physics can never be
reduced to geometry as Descartes had hoped.'' He came to this conclusion
because the failure of his theory had persuaded him that physics did not 
respect the symmetries, like scaling, that one would expect it to inherit
from space. He attributed the failure to the effect of quantum mechanics.
As will be explained in this and the companion paper \cite{ABFO}, scale-invariant
gravity exploits the symmetries that so appealed to Weyl's intuition, but does so
without compensating fields. Its realization in particle dynamics
in this paper uses no concepts that, except for the Newtonian notion of point masses,
 go beyond those of the three-dimensional
geometry axiomatized by Euclid. The geometrodynamical realization in \cite{ABFO} uses
nothing but the concepts of modern three-dimensional differential geometry, above
all Riemannian 3-spaces and scalar and 3-vector fields defined on them.

The key insight into scale-invariant gravity is the realization that the
apparent breaking of scale invariance might not be due to quantum
mechanics but to \textit{an inadequate theory of inertia}. As currently
described, inertia violates scale invariance. This fact seems to
have escaped notice. Moreover, the defect can be remedied without
compensating fields. Our approach to scale invariance is direct
and, we believe, new. It is universal because it can be implemented for any spatial geometry that admits a similarity
(scaling) symmetry.

The claim that the existing descriptions of inertia break scale
invariance develops a criterion taken from Poincar\'{e}'s
discussion of absolute and relative motion in \cite{Poincare1}. It
demands that, for any given spatial geometry, dynamics must be
\textit{maximally predictive} in a well-defined sense.
Relationists like Mach \cite{Mach} had argued that particle
dynamics should be expressed solely using quantities invariant
under Euclidean translations and rotations: the mutual separations
$r_{ij}$ of the particles. But Poincar\'{e} pointed out that this
can always be done through an automatic process of elimination. 
The real test is whether specification at an
initial time of the $r_{ij}$ and their time derivatives
$\dot{r}_{ij}$ (together with the particle masses and the force
law) suffices to predict the future evolution (of an isolated
system). Newtonian theory fails this precisely formulated
prediction test because the $r_{ij}$ and $\dot{r}_{ij}$ contain no
information about the total angular momentum \textbf{J}, and very
different evolutions result for different values of \textbf{J}. As a result,
when Newton's equations are expressed in terms of the $r_{ij}$ they contain
third derivatives with respect to the time. Clearly, such a theory is less
predictive than one containing only second derivatives. For more details, 
see \cite{Barbour2001}.

Poincar\'{e} did not attempt to formulate a maximally predictive
dynamics. Like Weyl later in the case of scaling, he simply
accepted that nature did not respect mathematical intuition. This
resignation may have been premature. Maximal predictive strength
is a powerful constructive principle.\footnote{In Appendix II of \cite{Einstein1955}
Einstein develops a notion of the `strength' of dynamical theories
 that bears some relation to Poincar\'{e}'s criterion. However, it
 is formulated in spacetime rather than configuration-space terms.} 
 It imposes very characteristic structures on theories that possess the property. The papers \cite{BOF, AB}
have shown how it can explain some very basic known laws of
nature: the universal light cone, the gauge principle, and Einstein's gravitational field equations. 
Now we use it in the hope of discovering \textit{new
physics}.

The first step is to push the principle to its limit. For
Poincar\'{e} still allowed Newton's absolute time and length
scale. Let us introduce configuration spaces. Let T be the
one-dimensional space of Newtonian absolute times, and Q be the
$3N$-dimensional Newtonian configuration space of $N$ point
particles of masses $m_{i}$. Let $\textrm{Q}_{\textrm{\scriptsize
RCS}}$ (RCS stands for \textit{relative configuration space}) be
the $(3N-6)$-dimensional space obtained from Q by quotienting with
respect to the Euclidean translations and rotations. Thus, a point
in $\textrm{Q}_{\textrm{\scriptsize RCS}}$ is the entire
6-dimensional group orbit in Q generated by them from a point
$q\in Q$. All particle configurations that can be carried into
each other by translations and rotations are identified. Finally,
quotienting by the dilatations, we arrive at the
$(3N-7)$-dimensional space $\textrm{Q}_{0}$ of shapes of
$N$-particle configurations: \textit{shape space}.

Standard Newtonian theory is formulated in the space
$\textrm{Q}\times\textrm{T}$. Given the masses and the force law,
its equations predict the future given an initial point $qt$ in
$\textrm{Q}\times\textrm{T}$ and initial direction in
$\textrm{Q}\times\textrm{T}$ at $qt$. (The direction in Q
determines the direction of the momenta, and the slope in T their
magnitude, i.e., the particle speeds.) Poincar\'{e}'s criterion is
met in $\textrm{Q}\times\textrm{T}$ but not in
$\textrm{Q}_{\textrm{\scriptsize RCS}}\times\textrm{T}$ and, a
fortiori, not in $\textrm{Q}_{0}$.

There are two known ways of constructing a maximally predictive
particle dynamics. The first uses directly the particle
separations $r_{ij}$ or, in a scale-invariant theory, ratios of
them. Schr\"{o}dinger \cite{Schrodinger} was one of many who
explored this route in the 20th century, but it leads to
anisotropic effective masses that are ruled out by sensitive
experiments \cite{HD}. The absence of mass anisotropy is a very tough
experimental test of relational theories. In \cite{BB}, Bertotti and I proposed a
maximally predictive form of variational dynamics that is
appropriately called \textit{best matching}. This method leads to
isotropic masses and is presented in section 4. In particle dynamics,
it leads to a small subset of the solutions allowed by Newton's
equations. The conditions that select them show how elimination of
non-relational kinematics leads to a dynamics that is fully
determined by its underlying geometry. In \cite{ABFO}, it leads to
a scale-invariant generalization of GR: \textit{conformal
gravity}.

Each symmetry has its own distinctive best matching and imposes a
constraint on the canonical momenta $\textbf{p}_{i}$. Simultaneously,
the best matching imposes further conditions on the Lagrangian and even on the solutions
(in conformal gravity) that ensure propagation of the constraint.
 It is illuminating to exhibit the
constraints and list the additional conditions.

Translational symmetry constrains the total momentum (all sums are
over $i=1-N$):
\begin{equation}
    \textbf{P}=\sum\textbf{p}_{i}=0,
\label{MC}
\end{equation}
and the potential $U$
must be translational invariant, so Newton's third law is enforced. 
These are not new restrictions,
since Galilean invariance enforces the invariance and ensures
(\ref{MC}) holds in the centre-of-mass (c.m) frame.

Rotational best matching leads to the non-trivial vanishing of the
total angular momentum $\textbf{J}$:

\begin{equation}
    \textbf{J}=\sum\textbf{x}_{i}
    \times{\textbf{p}_{i}}=0,
\label{JC}
\end{equation}
and in conjunction with translational best matching enforces $U=U(r_{ij})$. 
At this stage, the
energy $E$ is arbitrary, and inertial motion, for which
$U=\textrm{constant}$, is possible. The results (\ref{MC}) and
(\ref{JC}) were obtained in \cite{BB}. The unique (vanishing) value of $\textbf{J}$ imposed by best matching
eliminates the ambiguity of prediction noted by Poincar\'e.

The extension of best matching to dilatations is new and leads to
the most interesting result. It concerns the scalar quantity
$ D=\sum{\textbf{p}_{i}}\cdot{\textbf{x}_{i}}$, which has not, so far as I know, been given a name in the literature. Since it bears the same relation to expansion as angular momentum does to rotation (and has the same dimensions -- action), I shall call it the \textit{dilatational momentum}. Dilatational best matching forces it to vanish:
\begin{equation}
    \ D=\sum{\textbf{p}_{i}}\cdot{\textbf{x}_{i}}=0.
\label{VC}
\end{equation}
In addition, $U$ must be homogeneous of
degree -2 in $r_{ij}$, and $E$ must be zero. Thus pure inertial
motion, for which $U=\textrm{constant}$ and $E>0$, is not allowed.
It is in this sense that inertia violates scaling. \textit{There
is no maximally predictive inertial dynamics on shape space}. One cannot
formulate a theory of pure inertial motion without introducing additional
kinematic structure -- an absolute scale of length -- that mathematical intuition 
suggests one should not employ.\footnote{The situation in Newtonian mechanics is somewhat enigmatic. No absolute 
length itself can be observed (the overall scale is nominal),
but the effect of \textit{change of length} is observable. This matches what happens 
with regard to rotations. Newtonian mechanics is invariant under a time-independent rotation 
but not under changing rotations.}
 If one wishes to have any dynamics at all on 
shape space that satisfies the Poincar\'{e} criterion, \textit{it must include forces 
and have vanishing energy}. This casts new light on the old problem highlighted by
Hertz \cite{Hertz, Lanczos}: why does energy exist in two disparate and 
independent forms, kinetic and potential? Their separate existence is the reason why pure inertial motion is allowed in
Newtonian mechanics. In scale-invariant dynamics, this cannot be. The two forms of energy 
must come as inseparable twins and their sum must be zero. If the universe is scale invariant, this
has implications for dimensional analysis and our understanding
of the constants of nature (for a review of the present understanding, see \cite{Barrow}). In fact,
it leads to \textit{determination of the gravitational constant} G (Secs. 6 and 7).

So far as I know, formulating a scale-invariant 
theory through a requirement on the form of the initial-value
problem in shape space has not hitherto been considered. There is no 
indication of there having been work done in this direction in the 
often-cited review \cite{Fulton}. To avoid confusion, the reader should keep in mind this
new criterion of scale invariance (or conformal invariance). It is used
throughout both this paper and the companion \cite{ABFO}. 

Whereas $\textbf J$, like $\textbf M$, is conserved for isolated
Newtonian systems, nothing forces $\textbf J=0$. Moreover, the dilatational momentum $D$ is
not even conserved in general let alone zero. The best-matching
conditions are therefore non-trivial, and (\ref{VC}) is decidedly
restrictive. Note also that (\ref{MC})--(\ref{VC}) and the
associated conditions on $U$ and $E$ are as purely geometrical in
origin as the operators grad, curl and div of vector analysis, to
which they exactly correspond. This reflects the universal geometrical nature of
best matching.

It might seem that dilatational best matching is academic, since
gravity and electrostatics have $1/r$ potentials and not the
$1/r^{2}$ potentials required by (\ref{VC}). However, there exists
a unique, natural and universal way to select $1/r^{2}$ potentials
that are manifested as effective $1/r$ potentials
indistinguishable from the Newtonian counterparts (section 5). The strong equivalence 
principle is then satisfied for all these forces, including gravity. The
sole result of this procedure is to introduce a weak long-range
force $\textbf F$ like the one generated by Einstein's
cosmological constant $\Lambda$. However, unlike $\Lambda$, the
strength of $\textbf F$ is uniquely determined by the potential of 
the system and is therefore epoch dependent. Moreover, $\textbf
F$ forces the `size' of the universe, measured by the moment of
inertia $I$ (and by the spatial volume $V$ in conformal gravity),
to remain exactly constant. We shall see that there is a sense in 
which the cosmological force introduces an weak violation
of the strong equivalence principle. It is, however, a violation
that is reduced to the absolute minimum that is possible.

Another important thing to note is that the entire treatment of this
and the companion paper \cite{ABFO} is based on the premise that the 
universe can be treated as a self-contained closed dynamical system.
In the particle model, the moment of inertia of an $N$-particle system taken
to represent an `island universe' in Euclidean space plays an essential
physical role in the theory. In conformal gravity \cite {ABFO}, a similar
role is played by the 3-volume of the universe. Scale invariance that 
respects the SEP relies on the use of 
ratios of local separations divided by global quantities: the moment of inertia and the 
3-volume, respectively. Thus, the assumption of a self-contained universe is essential. 

Section 2 shows how the condition (\ref{VC}) can arise as a very
exceptional case within Newtonian theory and by how much Newtonian
theory fails to be maximal predictive. Section 3 shows how time is
eliminated by Jacobi's principle for the dynamical orbit of a
system in its configuration space. (Dynamical orbits should not be
confused with group orbits. Both exist in configuration space, and
both play important roles in best matching.) 
Section 4 introduces
best matching and develops the necessary formal techniques in Lagrangian form. 
Section 5 formulates the rule for passing from an arbitrary set of
Newtonian potentials to scale-invariant counterparts with the
associated cosmological force $\textbf F$. Section 6 considers how dimensional analysis is changed
by the elimination of time
and an absolute scale of length. It also shows how scale invariance leads to an explicit expression
for the gravitational constant G and the strength of \textbf{F}. Section 7 contains
estimates of the strength of $\textbf F$ predicted by
scale-invariant gravity and shows how the explicit expression for G in terms of the matter distribution of the universe permits a determination of the mass and size of the universe from the empirical value of G. Section 8 considers whether the actual
universe is scale invariant. The greatest need is for an
explanation of the Hubble red shift that does not rely on
expansion of the universe. It is this that makes the theory vulnerable to experimental disproof. Section 9 considers possible
quantum implications. Finally, the Appendix gives the Hamiltonian form of the theory. 

\section{The Lagrange--Jacobi Relation}

In the c.m frame, let an isolated system have Lagrangian
${\cal{L}}=T-U$ with
$T={\sum}{(m_{i}/2)}{\dot{\textbf{x}}_{i}}\cdot{\dot{\textbf{x}}_{i}}$
and potential $U$. The energy $E=T+U$ is conserved. The c.m moment
of inertia $I$ is
\begin{equation}
    I=\sum m_{i}\textbf{x}_{i}\cdot\textbf{x}_{i}\equiv
    {1\over
    M}\sum_{i<j}m_{i}m_{j}r_{ij}^{2},\hspace{.5cm}M=\sum{m_{i}}.
\label{MofI}
\end{equation}
Lagrange and Jacobi noted\footnote{I understand from $N$-body
specialists \cite{LJR} that the relation (\ref{Iddh}) of this section
is generally known as the Lagrange--Jacobi relation. For a useful account of
Lagrange's work, see \cite{Dziobek}} that $
\ddot{I}=2{\sum}m_{i}{\dot{\textbf{x}}_{i}}\cdot{\dot{\textbf{x}}_{i}}
+2{\sum}m_{i}{\textbf{x}_{i}}\cdot{\ddot{\textbf{x}}_{i}}$ has
important properties. By Newton's second law
$m_{i}\ddot{\textbf{x}}_{i}=-\partial{U}/\partial\textbf{x}_{i}$
and the definition of $T$, we find $
\ddot{I}=4T-2\sum\textbf{x}_{i}\cdot{\partial U\over\partial
\textbf{x}_{i}}$. If now $U$ is homogeneous of degree $k$, then,
using Euler's theorem and $T=E-U$, we find
\begin{equation}
    \ddot{I}=4(E-U)-2kU.
\label{Iddh}
\end{equation}
Consider Newtonian celestial
mechanics, for which $k=-1$. Then $\ddot{I}=4E-2U$, from which Lagrange deduced the
first qualitative result in dynamics. Since $U<0$ for gravity, $E\geq 0$ implies $ \ddot I>0.$ Thus $I$ is concave
upwards and must tend to infinity as $t\longrightarrow +\infty$
and $t\longrightarrow -\infty$. In turn, this means that at least
one of the interparticle distances must increase unboundedly, so
that any system with $E\geq 0$ is unstable.

Another consequence of (\ref{Iddh}) is the virial theorem. For
suppose that the system has virialized, so that $I\approx0$. Then
$4E=(2k+4)U.$

For our purposes, the most interesting consequence of (\ref{Iddh})
arises when $k=-2$. For then
\begin{equation}
    \ddot I=4E.
\label{Ih-2}
\end{equation}

Thus, $I$ has the parabolic dependence $I=2Et^{2}+bt+c$ on the
time and will tend rapidly to zero or infinity. Such a system is
extremely unstable, either imploding or exploding.

However, suppose $E=0$. Then $\ddot I=0$ by (\ref{Ih-2}), so that
$\dot I=2\sum m_{i}\dot\textbf x_{i}\cdot \textbf x_{i}=2\sum
\textbf p_{i}\cdot \textbf x_{i}=\textrm{constant}.$ Thus, the
dilatational momentum $D=\sum \textbf p_{i}\cdot \textbf x_{i}$ is conserved if
both $U$ is homogeneous of degree -2 and $E=0$. (The arbitrary
additive constant in $E$ is fixed by requiring $U\longrightarrow
0$ when all $r_{ij}\longrightarrow 0$.) Obtained thus,
conservation of $D$ is a fluke. However, as we just noted,
dilatational best matching requires not merely conservation but
vanishing of $D$. The homogeneity of $U$ and vanishing of $E$ are
enforced by the symmetry. Homogeneity is crucial for scale
invariance.

Newtonian dynamics has always been seen as the paradigm of
rationality. However, Poincar\'{e}'s analysis \cite{Poincare1} of
the initial-value problem shows that rationality is `arena
dependent'. If one treats the universe as an isolated dynamical
system in shape space, Newtonian dynamics needs more initial data
than seem necessary. It is worth spelling this out for the
three-body problem, for which shape space Q$_{0}$ is the
two-dimensional space of triangle shapes. Let it be coordinatized
by two angles $\alpha$ and $\beta$ of the triangle. If they are
taken as dependent and independent variables, respectively, the
orbit in Q$_{0}$ for a theory that satisfies the Poincar\'{e}
criterion should be described by
\begin{equation}
    {\textrm{d}^{2}\alpha\over
    \textrm{d}\beta^{2}}=f\left(\alpha,\beta,{\textrm{d}\alpha\over
    \textrm{d}\beta}\right),
\label{Dziobek}
\end{equation}
and three initial data would be
needed: $\alpha,\beta,{\textrm{d}\alpha/\textrm{d}\beta}.$

Newtonian theory fails this ideal of a \textit{dynamics of pure
shape}. To determine the curve of a generic solution in Q$\times$T
projected down to Q$_{0}$, \textit{five} further data are needed.
Two fix the direction of $\textbf J$ relative to the instantaneous
triangle, one is the constant ratio (rotational K.E/total K.E),
another is the varying ratio $T/U$, and the final one is the
varying ratio (K.E in change of size/K.E in change of shape).
There is a similar mismatch for the $N$-body problem. All the
extra data look natural in the Newtonian arena but incongruous in
shape space. Rotational best matching eliminates the first three,
dilatational best matching the other two.

\section{Jacobi's Principle}
The first step to a dynamics of pure shape is the elimination of
time by Jacobi's principle \cite{Lanczos}, which describes all 
Newtonian motions of one $E$ as
geodesics on configuration space. This topic has already been discussed in \cite{BOF, CQG94, EOT}, so
the ground already covered there will be only briefly recapitulated. However, this section will also consider 
the important issue of dimensions
in a timeless and scale-invariant theory, which has not hitherto been discussed. 

For $N$ particles of masses 
$m_{i}$ with potential
$U(\textbf{x}_{1}, \dots , \textbf{x}_{N})$ and energy $E$, the
Jacobi action is \cite{Lanczos}
\begin{equation}
    I_{\scriptsize\textrm{Jacobi}} =    2\int\sqrt{E -
    U}\sqrt{\tilde T} \textrm{d}\lambda,
\label{Jacobi}
\end{equation}
where $\lambda$ labels the points on trial curves and $ \tilde{T}
= \sum {m_{i} \over
2}{\textrm{d}{\textbf{x}}_{i}\over\textrm{d}\lambda}\cdot
{\textrm{d}{\textbf{x}}_{i}\over\textrm{d}\lambda}$ is the
parametrized kinetic energy. The action (\ref{Jacobi}) is timeless
since the label $\lambda$ could be omitted and the mere
displacements $d\textbf{x}_{i}$ employed, as is reflected in the
invariance of $I_\textrm{\scriptsize{Jacobi}}$ under the
reparametrization
\begin{equation}
    \lambda \rightarrow f(\lambda).\label{rep}
\end{equation}

In fact, it is much more illuminating to write the Jacobi action in the form
\begin{equation}
    I_{\scriptsize\textrm{Jacobi}} =    2\int\sqrt{E -
    U}\sqrt{T^*},\hspace{.5cm}T^*=\sum {m_{i} \over
2}\textrm{d}{\textbf{x}}_{i}\cdot
{\textrm{d}{\textbf{x}}_{i}},
\label{Jacobi*}
\end{equation}
which makes its timeless nature obvious and dispenses with the label $\lambda$.

The characteristic square roots of $I_{\scriptsize\textrm{Jacobi}}$
fix the structure of the canonical momenta:
\begin{equation}
    \textbf{p}_{i} =
    {\partial {\cal L} \over \partial(\textrm{d}\textbf{x}_{i}/\textrm{d}\lambda)} =
    m_{i}\sqrt{E - U \over {\tilde T}} {d \textbf{x}_{i}
    \over d \lambda},
\label{CanMom}
\end{equation}
which, being homogeneous of degree zero in the velocities, satisfy
the constraint \cite{Dirac}
\begin{equation}
    \sum{{{\textbf{p}}_{i}}\cdot
    {{\textbf{p}}_{i}}\over 2m_{i}} = E -
    U.
\label{QuadCon}
\end{equation}

The Euler--Lagrange equations are
\begin{equation}
    {\textrm{d}\textbf{p}^{i} \over \textrm{d} \lambda}= {\partial
    {\cal L} \over \partial \textbf{x}_{i}} = -\sqrt{{\tilde T} \over
    E - U}{\partial U\over
    \partial \textbf{x}_{i}} ,
\label{JacobiEL}
\end{equation}
where $\lambda$ is still arbitrary. If we choose it such that
\begin{equation}
    {{\tilde T} \over E - U} = 1 \Rightarrow {\tilde T} = E - U
\label{EnCon}
\end{equation}
then (\ref{CanMom}) and (\ref{JacobiEL}) become
$$
    \textbf{p}_{i} = m_{i}{{\textrm{d}\textbf{x}_{i}} \over \textrm{d}
    \lambda},\hspace{1.0cm} {\textrm{d} \textbf{p}_{i} \over
    \textrm{d}\lambda} = -{\partial U\over\partial \textbf{x}_{i}},
$$
and we recover Newton's second law w.r.t this special $\lambda$.
However, (\ref{EnCon}), which is usually taken to express energy
conservation, becomes the \textit{definition of time}. Indeed,
this emergent time, chosen to make the equations of motion take
their simplest form \cite{Poincare2}, is the astronomers'
operational ephemeris time \cite{Clemence}. It is helpful to see how `change creates
time'. The increment $\delta t$ generated by displacements
$\delta \textbf{x}_{i}$ is
\begin{equation}
    \delta t={\sqrt{\sum
    m_{i}\delta\textbf{x}_{i}\cdot\delta\textbf{x}_{i}}\over\sqrt{2(E-U)}}
    \equiv{\delta s\over\sqrt{2(E-U)}}.
\label{clem}
\end{equation}
Each particle `advances time' in
proportion to the square root of its mass and to its displacement, the total
contribution $\delta s$ being weighted by $\sqrt{2(E-U)}$. In a
scale-invariant theory, $E=0$, and $\delta t$ takes an especially
suggestive form: a given kinetic $\delta s$ `advances time' by an amount
inversely proportional to $|U|$.

\section{Best Matching}

Our goal is a theory of Jacobi-type geodesics on shape space
Q$_{0}$. To do this, we might interpolate shapes
$A_{0},B_{0}\in{\textrm{Q}_{0}}$ with suitably continuous curves
$q_{0}(\lambda)$ and extremalize an action of the form
$$
    {I_{\scriptsize\textrm{Q}}}_{0}=\int_{\lambda_{A_{0}}}^{\lambda_{B_{0}}}
    \textrm{d}\lambda
    \sqrt{f_{ij}{\textrm{d}q_{0}^{i}\over
    \textrm{d}\lambda}{\textrm{d}q_{0}^{j}\over
    \textrm{d}\lambda}},
$$
w.r.t certain shape coordinates
$q_{0}^{i}, \dots,q_{0}^{3N-7}$ on Q$_{0}$. The $f_{ij}$ will be
functions of the shape coordinates,
$f_{ij}=f_{ij}(q_{0}^{1},\dots,q_{0}^{3N-7})$, and the usual
summation is understood. But shape coordinates are awkward, and
this approach is impracticable.

Next, we try to work in Q and attempt an action of similar
form:
$$
    I_{\scriptsize\textrm{Q}}=\int_{\lambda_{A}}^{\lambda_{B}}
    \textrm{d}\lambda
    {\cal L},\hspace{.5cm}{\cal L}=\sqrt{f_{ij}{\textrm{d}q^{i}\over
    \textrm{d}\lambda}{\textrm{d}q^{j}\over
    \textrm{d}\lambda}},
$$
where now $i=1,\dots,3N$ and
$f_{ij}=f_{ij}(q)$. Since $I_{\scriptsize\textrm Q}$ should depend
only on the projected curve $q_{0}(\lambda)$ in Q$_{0}$, we seek
to make $\cal L$ invariant under $\lambda$-dependent
transformations that shift the $q(\lambda)$'s by arbitrary
$\lambda$-dependent amounts along their group orbits. The
$\lambda$-dependence is decisive.

Let us consider translations in one dimension and infinitesimally
differing $A$ and $B$. Then $\lambda_B-\lambda_A=\delta\lambda$
can be infinitesimal, and to first order we shall be minimizing
\begin{equation}
    I_{\textrm{\scriptsize{TS}}}=\sqrt{f_{ij}\delta
    q^{i}\delta q^{j}},\hspace{.5cm}\delta
    q^{i}={\textrm{d}q^{i}\over\textrm{d}\lambda}\delta\lambda,
\label{TS}
\end{equation}
where TS stands for thin sandwich. The action
(\ref{TS}) must be stationary to first order
under
\begin{equation}
    q^{i}(\lambda)\longrightarrow
    q^{i}(\lambda)+b(\lambda)e^i
\label{BM3}
\end{equation}
for \textit{all} choices of $b(\lambda)$. Here $e^i=1$ for all $i$ is the generator of translations for particle $i$.
If this is so, $I_{\scriptsize{\textrm{TS}}}$ will depend only on
the orbits of $A$ and $B$, since different choices of $b(\lambda)$
move $A$ and $B$ arbitrarily along their orbits. Note that under
(\ref{BM3}) the velocities change:
\begin{equation}
    v^{i}\longrightarrow
    v^{i}+b'e^i,\hspace{.5cm}v^{i}={\textrm{d}q^{i}\over
    \textrm{d}\lambda},\hspace{.5cm}b'={\textrm{d}b\over
    \textrm{d}\lambda}.
\label{BM4}
\end{equation}

Let us make a Taylor expansion of $b(\lambda)$:
$$b(\lambda)=b(0)+
b'(0)\delta\lambda+\dots.$$

The \textit{thin-sandwich problem} (TSP) requires us to minimize
(\ref{TS}) to first order w.r.t $b(0)$ and $b'(0)$. The
$\lambda$-dependence of the transformations is reflected in the
fact that for the TSP these are to be treated as independent
Lagrange multipliers. The first shifts $A$ and $B$ together along
their orbits; the second displaces one relative to the
other. This twofold freedom gives rise to a universal
characteristic structure of such variational problems. Provided
the TSP has a unique solution,\footnote{It will in the particle model, since positive-definite quantities
are being minimized; the situation remains open in
scale-invariant geometrodynamics \cite{ABFO}, which however has a well-behaved Hamiltonian formulation.} its limit as
$\delta\lambda\longrightarrow 0$ will clearly define a metric on
Q. For it can be used to define the `distance'
$I_{\textrm{\scriptsize{TS}}}$ (\ref{TS}) between any two
neighbouring points in Q$_{0}$. This process gets its name because
the `distance' is obtained when the configurations are
\textit{best matched}: brought as near to congruence, as measured
by (\ref{TS}), as possible.

To implement best matching in a convenient formal scheme, we
introduce an \textit{auxiliary gauge variable} $a\in\Gamma$, where
$\Gamma$ for one-dimensional translations is $R^1$. However, for a
general symmetry group defined by $d$ parameters, the auxiliary
space $\Gamma$ will have $d$ dimensions ($d=7$ for the full
Euclidean group). The space $\Gamma$ is used to extend the
configuration space Q to a larger formal space Q$\times\Gamma$.
\footnote{The auxiliary space $\Gamma$ is a fibre over the base Q,
but the formalism of fibre bundles is not needed for translations
and dilatations. It is for rotations, which are only mentioned in
this paper. For a beautiful gauge fibre-bundle treatment of
rotations in the $N$-body problem, see \cite{LR}.} Since Q
contains $d$ unphysical degrees of freedom, $d$ is the dimension
of the group orbits that foliate Q into Q$_{0}$ and is the
\textit{degeneracy}. Thus, the extension \textit{doubles the
degeneracy}. We shall denote the coordinates on $\Gamma$
collectively by $a$, which is a single variable in the present
case.

We now use the auxiliary $a$ to define the \textit{corrected
coordinate}\footnote{We shall use the words \textit{coordinate,
velocity, and momentum} in the sense of Hamiltonian canonical
theory.}
\begin{equation}
    \bar{q}^{i}=q^{i}+ae^i.
\label{CoCo}
\end{equation}

It is invariant under the pair of mutually compensating
transformations:
\begin{equation}
    q^{i}(\lambda)\longrightarrow
    q^{i}(\lambda)+b(\lambda)e^i,\hspace{.5cm}a(\lambda)\longrightarrow
    a(\lambda)-b(\lambda),
\label{banal}
\end{equation}
which, in view
of the triviality of the invariance thereby achieved, we shall
call the \textit{banal transformation}. It induces equally trivial
mutually compensating transformations of the velocities:

\begin{equation}
    v^{i}\longrightarrow v^{i}+b'e^i,
    \hspace{.5cm}a'\longrightarrow a'-b',\hspace{.5cm}a'={\textrm da\over\textrm d\lambda}.
\label{banalvelocities}
\end{equation}

We now define the \textit{corrected velocity} $\bar v^{i}$:
\begin{equation}
    \bar{v}^{i}=v^{i}+a'e^i.
\label{BM7}
\end{equation}

Unlike the \textit{naive (uncorrected) coordinate} $q^{i}$ and the
\textit{naive velocity} $v^{i}$, both $\bar{q}^{i}$ and
$\bar{v}^{i}$ are invariant under the banal transformation
(\ref{banal}). One correction compensates the other. It is
important to distinguish the transformation parameter $b$ from the auxiliary $a$.
Although they seem to `do the same thing' to the $q^i$'s [cf.
(\ref{BM3}) and (\ref{CoCo})], their conceptual status is quite
different.

Two rules now yield a fully invariant action of curves on Q$_{0}$:
1) On Q$\times\Gamma$, choose a Lagrangian $\bar{\cal L}=\bar{\cal
L}[\bar{q}^{i},\bar{v}^{i}]$ that is the square root of a
functional homogeneously quadratic in the $\bar{v}^{i}$ and
otherwise depends only on the $\bar q^i$; 2) ensure that the
action along any curve in Q$\times\Gamma$ is invariant w.r.t
\textit{arbitrary} variations $\delta a$ of $a$. The `arbitrary'
means that there are \textit{no fixed end points} of the $a$
variation. The point is that the $a$ variation is effectively
being used to solve successive TSPs to determine the metric on
Q$_{0}$. Were we to fix $a$ at the end points, we could not
determine the metric at them. We shall say that $a$ is varied by
the \textit{free-end-point method}. The $q$ variation is
different. It is being used to find the geodesics w.r.t the metric
determined by the $a$ variation.

Free-end-point variation has consequences of three kinds: 1) Conditions
on the $\bar{q}$ dependence of $\bar{\cal L}$; 2) constraints on
the canonical momenta that the Euler--Lagrange equations propagate
because $\bar{\cal L}$ satisfies conditions 1); 3) subsidiary
equations that must be satisfied in addition to the constraints
and the Euler--Lagrange equations. (This happens in GR and
conformal gravity \cite{ABFO}.)

The conditions on the $\bar q$ dependence of $\bar{\cal L}$
imposed under 1) can often cause the auxiliary $a$ to disappear
from $\bar{\cal L}$ (its velocity $a'$ never does). This happens for
translations, as we shall soon see, and rotations and in Abelian and non-Abelian gauge
theories.\footnote{A full treatment of this point goes beyond the scope of this paper. I hope to discuss it in a further paper devoted to the unifying perspective on gauge theory that best matching provides.} However, $a$ appears explicitly in scale-invariant
theories and also in a representation of GR \cite{ABFO}.

The reason for this is a characteristic property of scaling
transformations, which, in contrast to the additive nature of the
translations (\ref{BM3}), are multiplicative and change the norm. Indeed, the
compensating transformations for dilatations are
\begin{equation}
    q^{i}\longrightarrow bq^{i},\hspace{.5cm}a\longrightarrow {a\over
    b}.
\label{DilBanal}
\end{equation}
The corrected coordinate and
corrected velocity are
\begin{equation}
    \bar q^{i}=aq^{i},\hspace{.5cm}\bar v^{i}=av^{i}+q^{i}a'.
\label{DilCorr}
\end{equation}

The device of replacing the `partly redundant' $q^i$ by corrected
$\bar q^i$ is evidently universally applicable, i.e., it is always
possible to introduce a compensating transformation like the
second members of (\ref{banal}) and (\ref{DilBanal}) (which are
both `inverses' -- one by subtraction, the other by division).
More complicated symmetries and several simultaneous symmetries
are beyond the scope of this paper. However, translations and
dilatations in particle mechanics are a useful and adequate
preparation for the conformal symmetry in \cite{ABFO}. For simplicity, we take the Lagrangian
$\bar{\cal L}[\bar q^i,\bar v^i]$
diagonal in the $\bar v^i$:
\begin{equation}
    I_{\scriptsize\textrm{BM}} =\int\textrm{d}\lambda\bar{\cal L}[\bar
    q^i,\bar v^i]= \int{\textrm{d}}{\lambda}2\sqrt{B(\bar
    q^{1},\dots,\bar q^{N})T},\hspace{.5cm} T = {\scriptsize{1\over
    2}}{\sum_{i=1}^{N}}m_{i} (\bar
    v^{i})^{2}.
\label{Action}
\end{equation}

While we shall always insist on free-end-point variation w.r.t
$a$, the remaining Euler--Lagrange part of the variation can be
performed in two different ways. We can vary either w.r.t the
$q^i$ or w.r.t the $\bar q^i$. The latter variation is neater,
but we still use the former, since it shows how the absolute
part of Newtonian kinematics, i.e., the part not dictated by
three-dimensional Euclidean geometry, arises.

The physical variables $q^i$ and the auxiliary variable $a$ have the canonical momenta
$$
    p^{i}\equiv {{\partial\bar{\cal L}}\over\partial
    v^{i}}=\sqrt{B\over T}{\partial T\over\partial\bar v^i}{\partial\bar v^i\over\partial v^i},\hspace{.5cm}p^a\equiv {{\partial\bar{\cal L}}\over\partial
    a'}=\sum_i\sqrt{B\over T}{\partial T\over\partial\bar v^i}{\partial\bar v^i\over\partial a'}.
$$
Substituting the values of $\partial\bar v^i/\partial v^i$ and $\partial\bar v^i/\partial a'$ for translations and dilatations, we find
\begin{equation}
p^i_{\scriptsize\textrm t}=\sqrt{B\over T}{\partial T\over\partial\bar v^i},\hspace{.5cm}p^a_{\scriptsize\textrm t}=\sum_i\sqrt{B\over T}{\partial T\over\partial\bar v^i};\hspace{.7cm}p^i_{\scriptsize\textrm d}=\sqrt{B\over T}{\partial T\over\partial\bar v^i}a,\hspace{.5cm}p^a_{\scriptsize\textrm d}=\sum_i\sqrt{B\over T}{\partial T\over\partial\bar v^i}q^i.
\label{p}
\end{equation}

Therefore, in the respective cases we have the constraints
\begin{equation}
    \sum_i p^i_{\scriptsize\textrm t}\equiv
    P_{\scriptsize\textrm t}=p^a_{\scriptsize\textrm t},\hspace{.5cm}\sum_i p^i_{\scriptsize\textrm d}q^i\equiv D=ap^a_{\scriptsize\textrm d},
\label{Niall}
\end{equation}
which are primary \cite{Dirac} because they arise from the mere form of the action and not through variation (which gives rise to secondary constraints). The constraints (\ref{Niall}) reflect the banal invariance of the action. They are homogeneously linear in the $p^{i}$ with coefficients solely determined by the symmetry and arise because
$\bar{\cal L}$ depends on the corrected velocities $\bar v^i$. The actual form of
$\bar{\cal L}$ is immaterial. 

In accordance with the free-end-point method, the action must be stationary with respect to arbitrary variations of $a$. This means that at any instant both $\partial\bar{\cal L}/\partial a=0$ and $\partial\bar{\cal L}/\partial a'=0$. But $\partial\bar{\cal L}/\partial a'$ is the canonical momentum $p^a$ of the auxiliary variable, which must therefore vanish as a secondary constraint. The primary constraints (\ref{Niall}) now become
\begin{equation}
P_{\scriptsize\textrm t}=0,\hspace{.5cm} D=0,
\label{NiallBis}
\end{equation}
i.e., the momentum and dilatational momentum must vanish, as promised at the end of the Introduction.

As Dirac points out \cite{Dirac}, as soon as one has constraints in any dynamical system, one must verify that they are propagated by the Euler--Lagrange equations. An important feature of best matching is the manner in which its first condition $\partial\bar{\cal L}/\partial a=0$ automatically ensures propagation of the constraint enforced by
its second condition $\partial\bar{\cal L}/\partial a'=0$. We can see this in two ways.

First, since the action must be stationary for all variations of $a$, they will include variations with fixed end points, from which we deduce the standard Euler--Lagrange equation
$$
{\textrm dp^a\over\textrm d\lambda}={\partial\bar{\cal L}\over\partial a}.
$$

Since we also require ${\partial\bar{\cal L}/\partial a}=0$, this ensures that $p^a$ will be conserved. Thus, if $p^a$ is initially zero, it will remain zero. By the primary constraints (\ref{Niall}), the relations that hold for $p^a$ must also hold for the corresponding combinations $P_{\scriptsize\textrm t}$ and $D$ of the physical canonical momenta, so the constraints (\ref{NiallBis}) must propagate.

Second, we can, of course, confirm the propagation of (\ref{NiallBis}) directly, using the Euler--Lagrange equations of the physical variables in conjunction with the condition ${\partial\bar{\cal L}/\partial a}=0$. It is worth doing this to see the explicit form of this condition. In the case of translations
\begin{equation}
{\partial\bar{\cal L}\over\partial a}=\sum_i{\partial\bar{\cal L}\over\partial\bar q^i}{\partial\bar q^i\over\partial a}=\sum_i\sqrt{B\over T}{\partial B\over\partial\bar q^i}=0.
\label{q}
\end{equation}
From the Euler--Lagrange equations
$$
{\textrm dp^i\over\textrm d\lambda}=\sqrt{B\over T}{\partial B\over\partial\bar q^i},
$$
we directly conclude that
$$
{\textrm dP_{\scriptsize\textrm t}\over\textrm d\lambda}=\sum_i{\textrm dp^i\over\textrm d\lambda}=\sum_i\sqrt{B\over T}{\partial B\over\partial\bar q^i},
$$
which does indeed vanish by virtue of (\ref q). Of course, (\ref q) is the condition for the potential $B$ to be translationally invariant (which ensures fulfilment of Newton's third law). One of
 the easiest ways to achieve this is to choose
 $$
    B=B(|\bar q^i-\bar
    q^j|)=B(|q^i-q^j|),
 $$
 the second equation holding
 because the correcting $ae^i$ in the $\bar q^i$ simply drops out of the difference.
 Thus, for corrected coordinates formed
 additively the potential will not depend on the auxiliary gauge
 variable $a$. Also any constant $E$ can be added to the potential
 $B$, matching the total energy that appears in the Jacobi
 action (\ref{Jacobi}).

In the case of dilatations, we have a much more intimate interconnection of conditions. The condition 
$\partial\bar{\cal L}/\partial a=0$ gives
$$
0={\partial\bar{\cal L}\over\partial a}\equiv\sum_i\left(\sqrt{T\over B}{\partial B\over\partial\bar q^i}{\partial\bar q^i\over\partial a}+\sqrt{B\over T}{\partial T\over\partial\bar v^i}{\partial\bar v^i\over\partial a}\right)=\sum_i\left(\sqrt{T\over B}{\partial B\over\partial\bar q^i}q^i+\sqrt{B\over T}{\partial T\over\partial\bar v^i}v^i\right).
$$
We now divide by $a$ and add and subtract $Da'/a$, where $D$ is the dilatational constraint (dilatational momentum):
$$
0={1\over a}\sum_i\left(\sqrt{T\over B}{\partial B\over\partial\bar q^i}aq^i+\sqrt{B\over T}{\partial T\over\partial\bar v^i}(av^i+q^ia')\right)-{1\over a}\sum_i\sqrt{B\over T}{\partial T\over\partial\bar v^i}q^ia'.
$$
Since $av^i+q^ia'=\bar v^i$, $(\partial T/\partial\bar v^i)\bar v^i=2T$ and the constraint is required to vanish, we conclude
\begin{equation}
    \sum_i \left(\sqrt{T\over B}{\partial B\over\partial
    \bar q^i}\bar q^i+\sqrt{B\over T}2T\right)=0
    \Rightarrow\sum_i \left({\partial B\over\partial
    \bar q^i}\bar q^i+2B\right)=0.
\label{HomCond}
\end{equation} 
By Euler's theorem, this tells us
that the scheme will be consistent if $B$ is homogeneous of degree
-2 in the $\bar q^i$'s. The last step in establishing the mutual consistency of our interlocking system of equations is to note that the propagation of the dilatational constraint by the Euler--Lagrange equations leads to exactly the same condition (\ref{HomCond}).

To make the final connection with the
special Newtonian situation summarized at the end of section 1 and
derived from the Lagrange--Jacobi relation in section 2, we now
exploit the freedom to make a banal transformation and so pass
from a general frame of reference to the unique
\textit{distinguished representation} in which $a=1$ and $\lambda$ is
chosen to make $B/T=1$. In the distinguished representation, we recover
Newton's laws exactly subject to the homogeneity condition on the
potential and the dilatational constraint that the dilatational momentum
vanishes. The $q^i$ are then coordinates in an inertial frame of
reference, and the special $\lambda$ is indistinguishable from
Newton's absolute time. We have derived the non-geometrical
(absolute) part of Newtonian kinematics from a purely relational
scheme.

 As mentioned in the introduction, a similar treatment for the
 rotation group \cite{BB,LR} yields the constraint that the total c.m
 angular momentum vanishes (and a further restriction on the
 potential, which must be a function of the inter-particle
 separations). Note how well the relational origin of this
 dynamics is hidden. The constraints and the condition $E=0$
 apply to the complete `island universe' of $N$ particles in
 Euclidean space, and need not apply to subsystems like the solar
 system -- only the sums over all subsystems need satisfy the constraints.
 Since it is difficult to verify the constraints and the
 condition $E=0$ for the complete universe, the sole hard evidence for
 the relational origin of the dynamics is in the form of the
 observed potentials. The conditions imposed by the translations
 and rotations are respected by the observed Newtonian
 approximations to relativistic physics, but there seems to be a
 spectacular failure to match the condition needed for a fully
 scale-invariant dynamics. The next section will show how it may
 be only an apparent failure. However, let me first conclude this section with two comments.
 
 The first concerns the somewhat elaborate formalism of the free-end-point method. Why is it necessary? The velocity of our auxiliary variable plays exactly the same role as the scalar potential $\phi$ in the Hamiltonian formulation of Maxwell theory. Since $\phi$ has the dimensions of velocity and the action contains no variable of which it is the velocity, it should strictly be treated as an ignorable coordinate \cite{Lanczos}. It is, however, habitually treated as a multiplier, variation with respect to which yields the Gauss constraint, which is a stronger result than would follow from regarding $\phi$ as an ignorable coordinate. Our free-end-point method for translations shows why the multiplier treatment is correct. The Maxwell action contains the velocity of an auxiliary variable but not the variable itself. It would be possible to start with a more general action assumed to depend on both an auxiliary and its velocity. The variation with respect to the auxiliary would then lead to a condition on the Lagrangian that is, in fact, satisfied by the Maxwell Lagrangian, which, as in the case of particle translations, does not depend on the auxiliary after all. As already noted, this is the case in all gauge theories hitherto studied. However, as this and the companion paper \cite{ABFO} show, scale-invariant theories necessarily contain both an auxiliary variable and its velocity. The free-end-point method shows that correct results are obtained by treating both as independent multipliers even though one is the velocity of the other.

The second comment is closely related to the first and relates to Dirac's classic study of generalized Hamiltonian dynamics \cite{Dirac}. He presents the consistent propagation of constraints derived from singular Lagrangians as more or less a matter of luck, which it certainly is in some cases \cite{BOF}. However, as this paper and \cite{ABFO} show, best matching provides a transparent scheme in which constraints linear in the canonical momenta are automatically propagated because they reflect the existence of a geometrically defined metric on a quotient configuration space. The case of dilatational constraints provides a new and interesting example of consistent constraint propagation. Its generalization to geometrodynamics \cite{ABFO}, which closely follows the treatment given here, leads to decidedly nontrivial relations.
 
 \section{Scale-Invariant Dynamics with Specific Potentials}

We recall that the Jacobi action for a general Newtonian system is
$$
    I_\textrm{\scriptsize{Jacobi}} =    2\int\sqrt{E - U}\sqrt{\tilde T}
    \textrm{d}\lambda.
$$

Best matching with respect to translations and rotations restricts
the Newtonian possibilities to a relatively limited extent. The
total angular momentum must be zero and the potential $U$ must be
a function of the inter-particle separations. Best matching with
respect to dilatations has more far reaching consequences: the
energy $E$ must necessarily be exactly zero and the potential must
be homogeneous of degree -2.

We first show how Newtonian gravity can be recovered to a good
accuracy from a potential that is homogeneous of degree -2. There
are two easy ways to do this. Let
$$
    W=\sum_{i<j}{m_{i}m_{j}\over r_{ij}}.
$$
Thus, $W$ is minus the Newtonian gravitational potential with
Newton's constant G set equal to unity. The first `way to Newton'
is to take as potential
\begin{equation}
    U=-{W^{2}\over 2}.
\label{U1}
\end{equation}

In the distinguished frame with the special $\lambda=t$, the
equations of motion are
\begin{equation}
    {\textrm{d}\textbf{p}^{i}\over\textrm{d}t}=
    -W{\partial W\over\partial\textbf{x}^{i}}.
\label{FirstEoM}
\end{equation}

This is Newton's law with G replaced by $W$. If the system has
virialized and resembles a globular cluster, $W$ will remain
effectively constant, and the motions will be essentially
Newtonian. Astronomers will observe $1/r$ forces, but the hidden
dynamics will be scale invariant.

An island universe described by (\ref{FirstEoM}) can even mimic a
Hubble-type expansion even though expansion has no physical
meaning in scale-invariant dynamics. To see this, note that two
processes can change $W$. The first is general expansion or
contraction. Now this would change the moment of inertia
(\ref{MofI}),
$$
    I={1\over M}\sum_{i<j}m_{i}m_{j}r_{ij}^{2},
    \hspace{.5cm}M=\sum_{i}m_{i},
$$
but the vanishing of the dilatational momentum $D$ enforces constancy of $I$.
However, a changing shape of the matter distribution changes the
\textit{scale-invariant} Newtonian potential\footnote{The
scale-invariant potential in fact determines all the qualitative
properties of the Newtonian $N$-body problem and is a more important concept
than the ordinary potential. For example, all asymptotic limiting
motions of the 3-body problem terminate at one of the critical or
singular points of $U_0$. They `govern' the motion \cite{Marchal,
Rick}.}
\begin{equation}
    U_0=-M\sqrt{I}W.
\label{ScalInvPot}
\end{equation}
The constancy of
$I$ does not prevent the matter from `clumping', which will
increase both $U_0$ and $-W$, and with them the `gravitational
constant'. Suppose a planet orbiting a sun in an island universe
that is becoming clumpier. In the distinguished frame, the
gravitational forces will become stronger, and the planet--sun
separation will adiabatically decrease. However, if we insist
that, by definition, gravity is constant, this effect can be
offset by an adiabatic increase of all scales from the
distinguished frame. This will mimic a Hubble-type expansion.

In fact, although this is suggestive I doubt whether the Hubble red shift can be explained
this way. The analogue of squaring $W$ in the conformal setting
leads to a very complicated theory that may not even be consistent
\cite{ABFO}. However, there is the much more interesting possibility that uses the unique conserved quantity in a
scale-invariant universe, namely $I$, or rather $\sqrt{MI}=\mu$. This has a simple
conformal analogue, which is the volume of 3-space.

Therefore, to achieve scale invariance, and the strong equivalence principle,
we shall use
\begin{equation}
    \mu=\sqrt{\sum_{i<j}m_{i}m_{j}r_{ij}^{2}}.
\label{Rho}
\end{equation}

Just as one passes from special to general relativity (with gravity minimally coupled to matter) by replacing
ordinary derivatives in the matter Lagrangians by covariant
derivatives, Newtonian potentials can be converted into potentials
that respect scale invariance. One simply multiplies by an
appropriate power of $\mu$, which has the dimensions of length.
This is a rather obvious mechanism. What is perhaps unexpected is
that the modified potentials lead to forces $\textit{identical}$
to the originals accompanied by a universal cosmological force
with minute local effects. The scale invariance is hidden because
$\mu$ is conserved.

Let $U$ consist of a sum of potentials $U_{k}$ each homogeneous of
degree $k$:
\begin{equation}
    U=\sum_{k=-\infty}^{\infty}a_{k}U_{k}.
\label{NewtPot}
\end{equation}

The $a_{k}$ are freely disposable strength constants. The energy
$E$ in the Jacobi action (\ref{Jacobi}) will be treated as a
constant potential ($k=0$). (It plays a role like the cosmological
constant $\Lambda$ in GR).

Now replace (\ref{NewtPot}) by
\begin{equation}
    \tilde{U}=\sum_{k=-\infty}^{\infty}b_{k}U_{k}\mu^{-(2+k)}.
\label{ScaledPot}
\end{equation}

The equations of motion for (\ref{NewtPot}) are
$$
    {\textrm{d}\textbf{p}^{i}\over\textrm{d}t}=
    -\sum_{k=-\infty}^{\infty}a_{k}{\partial
    U_{k}\over\partial\textbf{x}^{i}};
$$
for (\ref{ScaledPot}) they are
$$
    {\textrm{d}\textbf{p}^{i}\over\textrm{d}t}=
    -\sum_{k=-\infty}^{\infty}b_{k}\mu^{-(2+k)}
    {\partial U_{k}\over\partial\textbf{x}^{i}}+
    \sum_{k=-\infty}^{\infty}(2+k)b_{k}\mu^{-(2+k)}U_{k}
    {1\over\mu}{\partial\mu\over\partial\textbf{x}^{i}}.
$$

Since $\mu$ is constant `on shell', we can define new strength
constants that are truly constant:
\begin{equation}
    b_{k}=a_{k}\mu^{2+k}.
\label{DefB}
\end{equation}
The equations for the modified potential become
\begin{equation}
    {\textrm{d}\textbf{p}^{i}\over\textrm{d}t}=
    -\sum_{k=-\infty}^{\infty}a_{k}
    {\partial U_{k}\over\partial\textbf{x}^{i}}+
    \sum_{k=-\infty}^{\infty}(2+k)a_{k}U_{k}
    {1\over\mu}{\partial\mu\over\partial\textbf{x}^{i}}.
\label{ModEq}
\end{equation}
If we define
\begin{equation}
    C(t)={\sum_{k=-\infty}^{\infty}(2+k)a_{k}U_{k}\over
    2\sum_{i<j}m_{i}m_{j}r_{ij}^{2}}
\label{DefC}
\end{equation}
and express $\mu$ in terms of $r_{ij}$, then equations
(\ref{ModEq}) become
\begin{equation}
    {\textrm{d}\textbf{p}^{i}\over\textrm{d}t}=
    -\sum_{k=-\infty}^{\infty}a_{k}
    {\partial U_{k}\over\partial\textbf{x}^{i}}+
    C(t)\sum_{j}m_{i}m_{j}{\partial{r_{ij}^{2}}\over\partial\textbf{x}^{i}}.
\label{AbbModEq}
\end{equation}

We recover the original forces exactly together with a universal
force. It has an epoch-dependent strength constant $C(t)$ and
gives rise to forces between all pairs of particles that, like
gravitational forces, are proportional to the inertial mass but
increase in strength linearly with the distance.\footnote{Although $C(t)$ is epoch dependent, this does not mean
that the theory contains any fundamental coupling constants with such a dependence. The epoch dependence is
an artefact of the decomposition of the forces into Newtonian-type forces and a residue, which is the
cosmological force.} The universal
force will be attractive or repulsive depending on the sign of
$C(t)$, which is an explicit function of the $r_{ij}$'s. For small
enough $r_{ij}$, they will be negligible compared with Newtonian
gravity. An estimate will be made in section 7, but a first conclusion
is this. A Newtonian-like universe with forces that appear to
violate scaling may nevertheless be scale invariant with an as yet
unrecognized cosmological force. Cosmologists might be greatly
deceived, but, having once got the idea, they could predict the
epoch-dependent $C(t)$ from the observed matter distribution and
the locally observed forces.

Moreover, their revised cosmological model will be stable: its
`size', measured by $\mu$, remains constant -- the automatic
adjustment of $C(t)$ ensures it. Scale-invariant cosmology is free
of the inescapable instability of Newtonian and Einsteinian
cosmologies. What is more, $C(t)$ is unambiguously fixed by a
fundamental symmetry, unlike Einstein's ad hoc $\Lambda$. It seems
paradoxical that a theory constructed to eliminate size as a
dynamical variable uses `size' ($\mu$) to generate a force that
conserves the size (in the distinguished representation). Of course, the
numerical value of the size is purely nominal. Moreover, the
paradox is an artefact of the formalism. Could we but use shape
coordinates conveniently, cosmology would be reduced to angles and
mass ratios.

Note also that the `size of the universe' is not kept constant
because a repulsive cosmological force balances attractive
gravity, as in Einstein's original cosmological model of 1917
\cite{Einstein17}. That achieved only unstable equilibrium.
Scale-invariant cosmology is stable whatever the sign of $C(t)$.
In Newtonian terms, the stability arises from the interplay of the
potential's homogeneity and the simultaneous vanishing of the
energy $E$ and the dilatational momentum $D$ (section 2). If one retains the same
forces but allows the minutest deviation from zero of $E$ or $D$,
the stability is lost, parabolically for $E\neq 0$ and linearly
for $D\neq 0$. This again shows how cosmologists could be tricked.
If they did not realize that a symmetry enforces stability, a
scale-invariant cosmology would appear to be balanced on a knife
edge. This is reminiscent of the `flatness knife edge' of modern
relativistic cosmology.

Before we obtain estimates, two potential flaws in the device of
achieving homogeneity with powers of $\mu$ should be mentioned. The first is due to the possibility of collisions, 
which lead
to infinities in the case of point particles. In
Newtonian $N$-body theory, the most intractable collisions are the
so-called central collisions in which all $N$ particles collide at
once \cite{Chenciner}. They cannot be regularized and are rather
like the Big-Bang or Big-Crunch singularities in GR. Both in Newtonian theory and
in GR, the cosmological singularities clearly stem from the
dynamical role of scale. Such singularities are not present in
scale-invariant cosmology \cite{ABFO}. However, lesser collisions
of $n<N$ particles are possible (and some form of gravitational
collapse may also occur in conformal gravity). In the particle
model, this will certainly cause $C(t)$ to become infinite even at
a two-particle collision, and the equations will be ill defined at
such collisions. There is a similar potential problem in conformal gravity, and further research will be needed to establish whether it is serious. The second possible difficulty is related to 
dimensional analysis, to which we now come.

\section{Dimensional Analysis}

In Newtonian dynamics -- and modern physics -- there are 
three fundamental physical dimensions: mass $[m]$, length $[l]$, and time
$[t]$. As long as no cosmological assumptions are made, these are necessarily
independent and absolute, in the sense that one believes that `they exist
out there in the world'. One is forced to choose local units to measure them.
If the universe is assumed to be self contained, the situation is different,
because then the local unit is merely a fraction of some global total. More significantly,
time can be eliminated as a dimension. One just has 
mass $[m]$ and length $[l]$. In scale-invariant geometrodynamics \cite{ABFO}, dimensional analysis is even 
simpler and is based on length alone.
The results obtained in this section are only the first step to this ideal. They are
nevertheless suggestive.  
The method used here  
develops further the analysis of \cite{BB1}.

In developing dimensional analysis, we consider two standpoints. The first is that of an imagined external observer
who possesses a god's eye view of the universe. The second is that of an internal observer forced to use locally 
chosen units of mass, length, and time. The internal observer has no knowledge of the universe as a whole. 
For both observers, length $[l]$ is the most important dimension. Indeed, time is measured by length -- the 
distance traversed
by the hand of a clock -- and masses are deduced from accelerations, which involve lengths and times. Let us therefore
assume both kinds of observers are equipped with rods, on which they mark units of length. These rods can be used to
measure the inter-particle separations $r_{ij}$. A measure of time $[t]$ is chosen by postulating that the motion of a 
particular body (the clock) measures time: when the clock traverses the already chosen unit of distance, one unit of
time elapses. Finally, as Mach showed \cite{Mach}, relative masses can be deduced from the reciprocal
ratio of the accelerations that bodies impart to each other when interacting in accordance with 
Newton's third law. If some given mass is chosen as the unit, other masses can then be determined in this way.

An extension of the same technique permits determination of charges. Since both masses and charges
are deduced from accelerations, it seems to me that they should all have the dimension mass, denoted $[m]$. 
We shall see that, at least for gravitational and electrostatic forces,
 this leads to a consistent scheme in which the action principle of the universe 
contains no dimensional coupling 
constants but only passive mass (Newton's original inertial mass) and active charges: gravitational (which happen to
be proportional to the passive mass), electrostatic and perhaps more. In standard dimensional analysis, 
the electric charge does not have the 
dimensions $[m]$ of mass, but it will in the proposed scheme. It then turns out (empirically) that the electric charge 
of the proton 
is $10^{19}$ times its gravitational charge (active gravitational mass), and both have dimension $[m]$. 
As we shall see at the end of this section, this assumption need not lead to conflict with
the fact that the fine-structure constant $\alpha=e^2/hc$ is dimensionless because Planck's constant 
$h$ and the speed of light $c$ will have different 
dimensions. 

Let us start by considering standard dimensional analysis in Newtonian theory for the case of gravitational and 
electrostatic interactions (of unit charges $|e_i|=e$). The Lagrangian is
\begin{equation}
    \sum_i m_i{\textrm{d}\textbf{x}_i\over\textrm{d}t}\cdot{\textrm{d}\textbf{x}_i\over\textrm{d}t}+
    \textrm{G}\sum_{i<j}{\bar m_i\bar m_j\over r_{ij}}-\sum_{i<j}{e_ie_j\over r_{ij}},\hspace{.5cm}|\bar e_i|=\bar e,
\end{equation}
where bars have been added to the masses in the gravitational potential to indicate that these are charges. Since the
dimensions of the kinetic energy are taken as fiducial and are $ml^2t^{-2}$, it is clear that Newton's constant G 
must have dimensions $m^{-1}l^3t^{-2}$ if the gravitational
charges are taken to have the dimensions of mass. It is equally clear that $e^2$ must have dimension $ml^3t^{-2}$.
This can be achieved by setting
\begin{equation}
    \sum_{i<j}{e_ie_j\over r_{ij}}=\textrm{G}\sum_{i<j}{\bar e_i\bar e_j\over r_{ij}},
\end{equation}
where $\bar e/\bar m$ is the ratio of the electrostatic force generated by unit electric charge to the gravitational force
generated by the unit mass $\bar m$. For definiteness, I shall assume that there are $N=2n$ particles of constant mass, 
$n$
of each carrying positive and negative unit electric charge. The total mass $M$ of the universe is equal to the sum of
the inertial masses and equal to the sum of the gravitational and electrostatic charges, the contribution of the latter
being in fact zero by the assumption of overall electric charge neutrality.

Let us now consider Jacobi's principle for Newtonian theory under the assumption that some hypothetical principle
forces the energy $E$ to be exactly zero. The action can then be written in the form
\begin{equation}
    A=2\sqrt \textrm G\int\sqrt{\sum_{i<j}{\bar m_i\bar m_j\over r_{ij}}-\sum_{i<j}{\bar e_i\bar e_j\over r_{ij}}}
    \sqrt{\sum_i {m_{i} \over
2}\textrm{d}{\textbf{x}}_{i}\cdot
{\textrm{d}{\textbf{x}}_{i}}},
\end{equation}
which has the normal dimensions $ml^2t^{-1}$ of action: $[\sqrt \textrm G]=m^{-1/2}l^{3/2}t^{-1}$ while
[integrand]=$m^{3/2}l^{1/2}$. However, the constant $\sqrt G$ in front of the action $A$, being a common factor, 
can have no effect on the observed motions and can be simply omitted. Then the action $\bar A=A$/G in Jacobi's principle 
has dimensions $m^{3/2}l^{1/2}$. If a non-vanishing energy $E$ is included, it will have dimensions $m^2l^{-1}$, 
as is evident from (\ref{Jacobi*}). To recover the normal dimensions of energy, we must take the Newtonian time
to have dimensions $m^{-1/2}l^{3/2}$. This is confirmed by the explicit expression (\ref{clem}) for the Newtonian time 
if, as above, we take the potential energy without G and therefore to have dimensions $m^2l^{-1}$.

Let us next consider the canonical momenta (\ref{CanMom}). They are homogeneous of degree zero in the velocities, and 
therefore any attempt to introduce a dimension of time into them will achieve nothing. They have the dimensions
$m^{3/2}l^{-1/2}$. It is true that if we multiply the dimensions by $m^{-1/2}l^{3/2}/m^{-1/2}l^{3/2}=m^{-1/2}l^{3/2}/t$, 
we recover the Newtonian
$mlt^{-1}$. However, this distorts the true nature of Jacobi geodesics, which involve only distances and masses.

We must now show how Newton's constant G is recovered as an emergent constant from the action $\bar A$. The Euler--
Lagrange equations that follow from this action are
\begin{equation}
    {\textrm{d}\over\textrm{d}\lambda}\left(\sqrt{\sum_{i<j}{\bar m_i\bar m_j\over r_{ij}}-
    \sum_{i<j}{\bar e_i\bar e_j\over r_{ij}}\over \sum_i {m_{i} \over
    2}{\textrm{d}{\textbf{x}}_{i}\over\textrm{d}\lambda}\cdot
    {\textrm{d}{\textbf{x}}_{i}\over\textrm{d}\lambda}}m_i{\textrm{d}\textbf{x}_i\over\textrm{d}\lambda}\right)
    =-\sqrt{\sum_i {m_i \over
    2}{\textrm d\textbf{x}_i\over\textrm d\lambda}\cdot
    {\textrm d\textbf{x}_i\over\textrm d\lambda}\over \sum_{i<j}{\bar m_i\bar m_j\over r_{ij}}-
    \sum_{i<j}{\bar e_i\bar e_j\over r_{ij}}}{\partial\over\partial\textbf{x}_i}
    \left(\sum_{i<j}{\bar m_i\bar m_j\over r_{ij}}-
    \sum_{i<j}{\bar e_i\bar e_j\over r_{ij}}\right)
\label{k}
\end{equation}

In the discussion of Jacobi's principle, we simplified this equation by choosing the unique $\lambda$ that makes
the square root on the right-hand side unity at all $\lambda$. In this way, we recovered Newtonian time. If this
method is applied to (\ref{k}), obtained from the action without G, the outcome can be interpreted in Newtonian
terms as due to a choice of the time unit that makes G=1. Clearly, 
if we make the square root equal to a constant $k$, we simply recover a different time that runs uniformly w.r.t
the first time but at a different rate. With this more general simplifying choice, (\ref{k}) becomes
\begin{equation}
m_i{\textrm{d}^2\textbf{x}_i\over\textrm{d}t^2}=-k^2{\partial\over\partial\textbf{x}_i}
    \left(\sum_{i<j}{\bar m_i\bar m_j\over r_{ij}}-
    \sum_{i<j}{\bar e_i\bar e_j\over r_{ij}}\right)
\label{GravEq}
\end{equation}
with `gravitational constant'
\begin{equation}
    k^2={\sum_i {m_i \over
    2}{\textrm d\textbf{x}_i\over\textrm d t}\cdot
    {\textrm d\textbf{x}_i\over\textrm d t}\over \sum_{i<j}{\bar m_i\bar m_j\over r_{ij}}-
    \sum_{i<j}{\bar e_i\bar e_j\over r_{ij}}},
\label{GravCon}
\end{equation}
which has the correct dimensions $m^{-1}l^3t^{-2}$. Let us now recall our internal observers within such a universe,
who have chosen arbitrarily units of mass, length, and time and use standard Newtonian dimensional analysis. 
Making purely local observations within their solar
system, they will find the the gravitational constant has some value G. Making cosmological observations, they
will discover that G has the value $k^2$ (\ref{GravCon}). Since they include G (=$k^2$) as a factor in the
potential energy, they will interpret (\ref{GravCon}) as saying that minus the Newtonian potential energy is
equal to the kinetic energy. Thus, they will find that the total energy of the universe is
exactly zero and will interpret this as a `cosmic coincidence' since they can find no theoretical argument that
enforces the condition.

From the point of view of a dynamics of pure shape, Newtonian theory in Jacobi form is clearly imperfect, this being
reflected in the fact that the action depends on the length scale. Also, no fundamental
principle justifies the assumption that $E=0$. However, the dependence on the mass scale is not serious,
because all masses are constant and can be expressed as dimensionless ratios of the total mass $M$. 

Let us now repeat the dimensional analysis and `calculation of G' for scale-invariant gravity (SIG) and 
electrostatics with the assumption that
the necessary homogeneity of degree -2 of the potential is achieved as above by means of $\mu$. This is a much more
interesting theory because both the energy and the dilatational momentum must be exactly zero. 

As is evident from (\ref{GravEq}), the basic form of the equation of motion for Jacobi-type actions (with $E=0$ either
by stipulation or enforced by scale invariance) is
$$
    m_i{\textrm{d}^2\textbf{x}_i\over\textrm{d}t^2}=-{T\over U}{\partial U\over\partial\textbf{x}_i}
$$
with $k^2=T/U$ identified as Newton's constant. To pass from Newtonian theory to scale-invariant gravity, we 
simply replace the Newtonian potential $U$ by $U/\mu$. Then the equation of motion becomes
\begin{equation}
    m_i{\textrm{d}^2\textbf{x}_i\over\textrm{d}t^2}=-{T\mu\over U}{\partial \over\partial\textbf{x}_i}
    \left({U\over\mu}\right)=-{T\over U}{\partial U\over\partial\textbf{x}_i}+
    {T\over \mu}{\partial \mu\over\partial\textbf{x}_i}.
\label{ScaleGravEq}
\end{equation}

We have the identical expression $T/U$ for G, but there is now a more subtle reason for its constancy. First,
$T\mu/U$ is made constant by choosing one of the distinguished times. This is the device that made $T/U$=G
constant in the Newtonian case. Second, $\mu$ is constant in scale-invariant gravity as a consequence of the
dynamics. Thus, in (\ref{ScaleGravEq}) the coefficient of the gravitational force is constant because of these 
two reasons together. The coefficient of the cosmological force is epoch dependent. Both coupling constants are 
emergent and uniquely determined. We shall estimate them in the next section. The strength of electrostatics 
relative to gravity is constant and, in the absence of unification of the forces of nature, must be determined
empirically.

Two possible problems in scale-invariant gravity were noted at the end of the previous section. The one to do 
with dimensional analysis concerns the extension of the method presented in this section to Newtonian forces with
other than $1/r$ distance dependence. Presumably, one will always want the dimensions of all terms in the potential
to be the same. For gravity and electrostatics, the dimensions are $ml^{-2}$, but it seems to me that it will be
difficult to find the right combination of charges to give these dimensions for other than $1/r$ Newtonian forces without
introducing undetermined constants with dimensions of mass. This may not be a problem or, if it is, it might even
be a virtue through ruling out such forces, which are not observed. However, I think the particle model is probably 
not a reliable guide for such arguments. At the least, there is no problem with gravity and electrostatics.

Continuing with the dimensional analysis, calculation of the various 
dimensions of the most important dynamical quantities for scale-invariant gravity (SIG) together with
the dimensions for standard Newtonian theory (SNT) and `Jacobianized' Newtonian theory (JNT) as calculated 
earlier gives the following results:
$$
    \hspace{1cm}Action\hspace{.5cm}Length\hspace{.5cm}Momenta\hspace{.5cm}Time\hspace{.5cm}Energy
$$
$$
    \textrm{SNT}\hspace{.3cm}ml^2l^{-1}\hspace{.7cm}l\hspace{1.4cm}mlt^{-1}
    \hspace{1.1cm}t\hspace{1.1cm}ml^2t^{-2}
$$
$$
    \textrm{JNT}\hspace{.4cm}m^{3/2}l^{1/2}\hspace{.8cm}l\hspace{1cm}m^{3/2}l^{-1/2}
    \hspace{.5cm}m^{-1/2}l^{3/2}\hspace{.5cm}m^2l^{-1}
$$
$$
    \textrm{SIG}\hspace{1cm}m\hspace{1.2cm}l\hspace{1.5cm}ml^{-1}
    \hspace{1.2cm}l^2\hspace{1.2cm}ml^{-2}
$$

The results for the scale-invariant case are striking and suggestive. First, since mass is a ratio, 
the action is effectively dimensionless. This suggests that for a scale-invariant theory the quantum of action
will be a pure number. A similar calculation for scale-invariant geometrodynamics also suggests that action
is a pure number. Also remarkable are the dimensions of the canonical momenta, which exactly mirror the de Broglie
relation for the momentum in quantum mechanics. It is worth emphasizing that these results are non-trivial
and arise from a `trinity' of requirements: no time, scale invariance, and the implementation of the 
strong equivalence principle by the use of $\mu$. 

The at first enigmatic result that in the scale-invariant case `time is an area' is very natural. One of 
the length dimensions arises from the 
fact that the hand of clock must traverse a length to measure time. This length is supplied by the particle displacements 
in the numerator of (\ref{clem}). The other length comes from the weighting by the denominator of (\ref{clem}) and the 
fact that scale invariance and the SEP require the potential to have dimensions $ml^{-2}$. 

Let us now consider what happens to the fine-structure constant by applying the rule for passing from the Newtonian 
dimensions $m, l, t$ to their scale-invariant timeless 
counterparts, which is simply to replace $t$ by $l^2$ and remember that the Newtonian potentials have been
divided by $\mu$. Thus, the Newtonian dimensions of velocity are $[v]=lt^{-1}$, but now $[v]=l^{-1}$. Similarly,
the Newtonian dimensions of Planck's constant are $[h]=ml^2t^{-1}$, but now they become $[h]=m$, as we have already 
found. Thus, the $hc$ that appears as the denominator of the fine-structure constant no longer has its
Newtonian dimensions $ml^3t^{-2}$ but $ml^{-1}$.
Let us now consider the electric charge $e$. In Newtonian mechanics, it appears in the potential 
$\psi=\sum_{i<j}e_ie_j/r_{ij}$,
but in the new scheme this is replaced by $\psi\mu^{-1}$. In Newtonian mechanics, $e^2/hc$ is dimensionless, so
$[e^2]=ml^3t^{-2}$. However, if our approach is on the right track, the quantity that hitherto has been taken to be
$e^2$ is actually $e^2\mu^{-1}$, and $[\mu]=ml$. Therefore, if we postulate $[e]=m$, the new dimensions of $e^2$ 
are $ml^{-1}$, which is 
what we just found for $[hc]$, so the `new fine-structure constant' is still dimensionless. Even though gravitational
charge does not appear to be quantized in the way electric charge is, one can still define a `gravitational
fine-structure constant', which will also be dimensionless.

Only if and when a scale-invariant quantum cosmology has been created will it be possible to say whether the relations 
found in this section have a deep meaning or are merely fortuitously suggestive. However, there must surely be
an \textit{a priori} presumption that scale invariance will overturn current ideas about the fundamental constants
of nature. When Planck \cite{Planck} deduced his famous units of length, time, and mass, he employed $h, c, G$. 
He noted that the units will have fundamental significance so long as the theories from which they are deduced
are correct. The analysis of this section suggests that, if the universe is self contained and scale invariant, 
the quantum of action is a pure number. If so, it seems inevitable that a reorientation of our ideas about dimensional
analysis and the constants of nature will be necessary.

\section{Estimates}

Let us make estimates for the model with gravity and electrostatics.
Suppose an island universe of $N$ point particles half of which
carry positive unit charges $e_{i}$ and the other half negative
charges $-e_{i}=-1$ ($N$ is an even integer). With the notation of the previous section, take the Jacobi-type
action to be
\begin{equation}
    I_{\scriptsize\textrm{Model}}=\int\textrm{d}\lambda\sqrt{{E\over\mu^{2}}+
    {1\over\mu}\sum_{i<j}{\bar m_{i}\bar m_{j}\over r_{ij}}-
    {1\over\mu}\sum_{i<j}{\bar e_{i}\bar e_{j}\over
    r_{ij}}}\sqrt T =\int \textrm{d}\lambda\sqrt P\sqrt T,
\label{ModelAction}
\end{equation}
where $T$ is best matched.

Although no principle forbids the `energy' term $E/\mu^{2}$ in the
potential, it will merely lead to a further cosmological force
that adds no interest and introduces the undetermined constant $E$ (with dimension $[m]$). 
Let us therefore omit it, and hope that
eventually some principle will rule it out -- and determine the
electrostatics-to-gravity strength ratio $\bar e/\bar m$, which for the moment must be fixed
empirically.  Let us
make estimates using $\bar e^2/\bar m^2\approx 10^{40}$,
reflecting the relative strengths of the electrostatic and
gravitational interactions for the common elementary particles. Let us
also take $N=10^{80}$ particles, since this is the estimated
number of nucleons in the observed universe within the Hubble
radius H. For simplicity, let us take all masses $\bar m_i=\bar m$ with $\bar mN=M$.

As cosmological model, we first assume a spherically symmetric and
more or less uniform `plasma' matter distribution like a globular
cluster with mean radius $R$. In
the estimation of (minus) the potentials
$$
    \mu,\hspace{.5cm}\psi=\sum_{i<j}{\bar m_i\bar m_j\over r_{ij}},
    \hspace{.5cm}\phi=-\sum_{i<j}{\bar e_{i}\bar e_{j}\over r_{ij}},
$$
the fact that all masses are positive while half the charges are
positive and half negative leads to big differences. Both $\mu$
and $\psi$ contain $\approx N^{2}$ terms formed by all pairs of
particles, for which the mean separation will be $R$. Each
mass `feels' all the other $N-1$ masses, and the mean
contribution of each pair will be $\bar m^2/R$. In contrast, each charge
$\bar e_{i}$ finds itself in an almost exactly neutral cloud. In fact,
the total charge outside any considered charge $\bar e_i$ will be
$-\bar e_{i}$, and the `mean distance' of the opposite charge will be
$R$. Therefore, to a good accuracy the electrostatic potential $\phi$ consists of $N$
terms, each of magnitude $\approx \bar e^2/R$.\footnote{Physicists wonder how it will ever be possible
to explain the huge difference between the strength of gravity and the remaining forces of nature.
It seems to me not impossible (see the first footnote on p. 23 of \cite{BB1}) that, in a self-contained universe, 
it is somehow a consequence of
the very different ways in which the potentials for gravity and electrostatics are calculated. 
This might have far-reaching consequences in quantum cosmology.} 
(The minus has disappeared because the
charges are opposite.)

Therefore, in the `plasma' state
\begin{equation}
    \mu\approx N\bar mR,\hspace{.5cm}\psi\approx{N^2\bar m^2\over R},
    \hspace{.5cm}\phi\approx{N\bar e^2\over R},
\label{Estimates}
\end{equation}
and the corresponding contributions to $P$
in (\ref{ModelAction}) are
\begin{equation}
    P_{\scriptsize\textrm{plasma}}={\psi\over\mu}+
    {\phi\over\mu}\approx
    {N\bar m\over R^{2}}+{10^{40}\over R^{2}},\hspace{.5cm}\bar m=1,
\label{PlasmaEst}
\end{equation}
so that for $N\sim 10^{80}$ the (positive) electrostatic
contribution to $P$ will be only $10^{-40}$ of gravity's.

Now suppose the `plasma' evolves into a `post-nucleosynthesis'
state in which all the charges are in $N/2$ neutral pairs with
inter-charge separation $\bar r$. This pairing will change $\phi$
to $\phi\approx -N\bar e^2/\bar r$, have negligible effect on $\mu$ and will
only change $\psi$ significantly if $\bar r\leq 10^{-80}R$. The
possibility of `gravitational collapse' to such an extremely small
relative size, which corresponds to $\sim
10^{-20}l_{\scriptsize\textrm{Planck}}$ for $R\sim
10^{28}\textrm{cm}$, will not be considered here. Instead, let us
consider the more immediately realistic case $\bar r/R\sim
10^{-40}$, so that $\bar r$ is of nuclear order
$10^{-12}\textrm{cm}$. Then, in contrast to the `plasma' estimate
(\ref{PlasmaEst}), we obtain the `post-nucleosynthesis' estimate
\begin{equation}
    P_{\scriptsize\textrm{pns}}={\psi\over\mu}-
    {\phi\over\mu}\approx
    {N\bar m\over R^{2}}+{10^{40}\over R\bar r},
\label{PNS}
\end{equation}
so now the two contributions will have equal orders of magnitude.

Making the assumption that the particle model can give valid order-of-magnitude estimates, let us now try to
estimate the size and mass of the actual universe. In accordance with the above estimates, let us assume that
the combined potential $U$ of the gravitational and electrostatic potentials is $\approx M^2/R$. Then
\begin{equation}
    G={T\over U}={R\sum m_i\sum_i{\textrm d\textbf{x}_i\over\textrm d t}\cdot{\textrm d\textbf{x}_i\over\textrm d t}
    \over M^2}.
\end{equation}

To extract useful information from this, it is necessary to identify the characteristic values of the velocities
that appear in the numerator. Since special relativity prohibits velocities above the speed of light $c$, while
relativistic speeds are realized in nuclei, let us assume for simplicity that all velocities are $\approx c$. Then
\begin{equation}
    G\approx {RMc^2\over M^2}\approx {c^2\over\rho R^2},
\end{equation}
where $\rho$ is the matter density, assumed constant. Since $G$ and $c$ are known accurately, and $\rho$ is known
approximately, we can estimate the `radius of the universe':
\begin{equation}
    R\approx\sqrt{c^2\over G\rho}.
\end{equation}

Taking $\rho\approx 10^{-30}\textrm{g}.\textrm{cm}^{-3}$, and substituting the known values of $c$ and $G$, we find
\begin{equation}
    R\approx 10^{29}\textrm{cm},\hspace{.5cm}M\approx 10^{57}\textrm g.
\label{Size}
\end{equation}

Such relationships are obtained in modern cosmology, in which they reflect the fact that the universe has a near-critical
density and is therefore spatially flat. The estimates for the radius and mass of the universe refer to the current
epoch and increase with time. Inflation is invoked to explain the flatness, which must otherwise be regarded
as a cosmic coincidence. In scale-invariant gravity, the 
interpretation is different. An analogue of the flatness coincidence is a prediction that follows from the basic
dynamical structure of the theory. Moreover, the radius $R$ of the universe cannot change significantly because
$\mu\approx MR$ is constant. Of course, $R$ has no absolute meaning. The significance of the above estimates is 
that they relate human dimensions to the cosmic dimensions. The issue of whether scale-invariant gravity has any 
chance of replacing the standard hot big-bang model will be discussed in the next section. In this section, it only
remains to estimate the relative strengths of gravity and the cosmological force.

Since
$\textbf F$ increases linearly with the distance $r$ while gravity
decreases as $1/r^2$, the relative strengths of the two forces
will depend on $r$. The simple calculation shows, first, that the cosmological force must be attractive if
the only Newtonian forces are gravity and electrostatics\footnote{The expressions (\ref{ModEq}) and (\ref{DefC})
simplify if there is only one value of $k$, and we can use the equation (\ref{ScaleGravEq}).} and, second, that at distance
$\tilde{r}$ the ratio of $\textbf F$ to
gravity is
\begin{equation}
    {\textrm{cosmological force}\over{\textrm{gravitational force}}}\sim
    {r^{3}\over R^{3}}.
\label{Ratio}
\end{equation}

This result can be expected to survive in conformal gravity. It
shows that on solar-system scales, for which $r/R\sim 10^{-14}$,
the effect of $\textbf F$ will be $\sim 10^{-42}$ and negligible.
However, for $r\sim R$, the two forces will have the same
magnitude. The present derivation of the cosmological force as a
scaling concomitant of gravity (and possibly other forces in
certain epochs) differs markedly from all ad hoc introductions of
a cosmological constant: $\textbf F$ must occur, its strength is
fixed and extremely weak within the solar system, and it is always
`tuned' to maintain a universe of constant size.\footnote{One can see qualitatively how this happens. Suppose a particle in a `cloud' that forms an island universe manages to escape a significant distance from the cloud. It will then feel a restoring force that grows linearly with distance and will be rapidly pulled back into the cloud as if it were on a piece of elastic. Similarly, if the cloud becomes egg shaped, the particles at the ends of the long axis will feel an enhanced restoring force.} The relative
strength of $\textbf F$ may increase in a `post-nucleosynthesis'
phase.

\section{An Alternative Cosmology?}

It seems barely possible that scale-invariant gravity could replace GR
and give a viable cosmology. The standard hot Big Bang scenario looks very secure.
However, scale invariance is an
attractive principle, and the manner in which GR \textit{just} fails to be scale invariant (spelled out
in \cite{ABFO}) is mysterious. Moreover, it is precisely this `failure' that makes an expanding universe possible. For this reason, the best-matching scheme, with its definite
and striking predictions (\ref{Size}) and (\ref{Ratio}), can, at the least, act as a useful foil to the
current paradigm.

In this spirit, I shall merely make a few comments.
Serious conclusions must await detailed calculations using
conformal gravity, the structure of which is now settled
\cite{ABFO}. However, it looks as if the first obvious test will be passed easily: as \cite{ABFO} shows, conformal
gravity at human, solar-system and binary-pulsar mass and distance
scales appears to agree with GR as well as the scale-invariant
(\ref{ModelAction}) agrees with Newtonian gravity. The real problems are all in cosmology, in which the major difference between the two theories is this. In the
standard model the universe is simultaneously expanding and becoming steadily more inhomogeneous.
In the scale-invariant case, the universe cannot expand, since that has no meaning. It can only become
more inhomogeneous -- it can only `change its shape'.

Now it is a fact that, while remaining isotropic and homogeneous on large scales, the presently observed universe is obviously hugely more structured than it was in the past. It has changed its shape massively. The estimates of section 7 show how readily the scale-invariant potential energy can increase if the universe becomes more clumpy. Scale-invariant gravity must, in the first place, yield a cause of the Hubble red shift. The only plausible candidate that I can see is this change in the `potential' of the universe induced by such clumping. It is suitably great and, according to the standard model, has been happening since the end of inflation. Therefore, the conjecture has to be that somehow the change in potential causes the Hubble red shift. This is not inherently impossible. We know that differences in the gravitational potential give rise to a gravitational red shift. Moreover, in scale-invariant gravity the increases in all forms of potential energy increase the cosmological force, which, being universal and acting on mass, is gravitational in nature. Therefore, it seems to me not entirely impossible that clumping could give rise to some kind of gravitational red shift. In the realm of classical physics, the integrated Sachs--Wolfe \cite{SW} effect would certainly be relatively more important than it is in Einsteinian cosmology. However, it seems to me that it must still fall far short of what is needed. Possibly more relevant is the fact that, in scale-invariant gravity, the cosmological force mimics a negative cosmological constant whose strength may have increased by orders of magnitude over cosmological epochs. Its impact on galactic dynamics is obviously a matter of some interest, as is the possibility that it might somehow have a bearing on the red shift. For the moment, I do not think that anything more can be usefully said about this crucial issue, except perhaps that scale-invariant gravity does force one to look for totally new explanations for fundamental effects. That cannot be bad. In the next section, I shall mention possible quantum effects induced by clumping. 

At this stage, it is premature to try to attack detailed issues such as primordial nucleosynthesis and the origin of the microwave background. However, the isotropy and homogeneity of the observed universe on large scales and the problem of singularities seem to me to much more promising topics. I shall discuss them in the context of particle dynamics.

In considering these issues, the difference between
the configuration spaces Q and Q$_0$ for theories with and without
scale invariance is significant. Consider the Newtonian
gravitational $N$-body problem for $N\geq 3$. It is dominated by
the uniquely distinguished point in Q at which
all particles coincide. In \cite {EOT} I have called this point (and its analogues in other configuration spaces) Alpha. The singularity of both the topology and
the potential at Alpha causes the unregularizable central
collisions mentioned earlier. But Alpha is not a `shape' and does
not belong to shape space Q$_0$ \cite{Kendall}. However, Q$_0$ has a
rather different uniquely distinguished configuration Alpha$_0$.
It is the most uniform state, at which the (negative)
scale-invariant potential has an absolute maximum. In the 3- and
4-body problems, this is at the equilateral triangle and
quadrilateral, respectively, whatever the masses of the particles \cite{Marchal}.
The topology, potential and scale-invariant dynamics are very well
behaved at Alpha$_0$. There cannot be anything like the Big-Bang singularity.

There is another difference. The dynamical curve of the Newtonian
$N$-body problem can enter (or leave) Alpha with arbitrary energy
$E$ and dilatational momentum $D$ (but only with vanishing angular momentum). If
emergence from Alpha is seen as a model Big Bang, the statistics
of initial states must include a spectrum of $E$ and $D$ values.
In contrast, $E=D=0$ in scale-invariant dynamics. Intuitively, the
fulfilment of the constraint requiring the dilatational momentum to vanish, especially in conformal
gravity, where it becomes a constraint at each space point, must
favour isotropic momenta distributions. At the very least, the undoubted
existence of the smooth absolute maximum of the scale-invariant
potential at the most uniform state must be significant. The
critical points of the potential always exert a decisive influence
on the classical \cite{Rick, Chenciner} and quantum \cite{Tanner} physical processes that unfold in a
configuration space. I believe scale invariance opens up new possibilities for explaining (without fine tuning and the need to postulate as yet unknown scalar fields) those features of the universe for which inflation is currently invoked.

As a final example, let me mention the flatness problem. The current theory is unable to explain why the universe seems to be expanding at almost exactly its escape velocity. In Newtonian terms, this corresponds to expansion with total energy $E$ exactly equal to zero. In the standard theory, there is no fundamental principle that can enforce $E=0$. However, we have seen that this is exactly what scale invariance does do. Of course, at the same time, it rules out expansion, predicting, in particle dynamics, a universe like a globular cluster. Nevertheless, there is a principle that enforces $E=0$.

To conclude this tentative discussion of classical cosmology, scale invariance encourages radical questioning of accepted ideas and opens up some interesting possibilities.

\section{Possible Quantum Implications}

The dimensional analysis discussed in section 6 indicates that if the universe is described by a 
scale-invariant law it will be necessary to revise our ideas about the constants of nature.
Section 6 shows that it is in principle possible to calculate Newton's constant G in a scale-invariant
framework and suggests that the quantum of action does not have dimensions $ml^2t^{-1}$ but is a pure number.
These major changes are a consequence of the elimination of time from kinematics, the transition to potentials
that are homogeneous of degree -2, and the inclusion of $\mu$, the square root of the moment of inertia of the
universe, in the fundamental action principle.

There is another very direct way of seeing that potentials homogeneous of degree -2 may well have far-reaching
consequences. The need for such potentials is a direct
consequence of having a kinetic energy quadratic in the
velocities. Each velocity has the dimension of length (divided by
time), and this is why we need the degree -2 in the potential. Now
the quadratic kinetic energy of classical physics leads to the
Laplacian in quantum physics. If one solves the time-independent
Schr$\ddot{\textrm o}$dinger equation for the $1/r$ electrostatic
potential, there is a mismatch between the dimensions of the
potential and the Laplacian. A constant with the dimensions of
length is needed. It seems to me that this is the origin of the
Planck length. A scale-invariant time-independent
Schr$\ddot{\textrm o}$dinger equation would need no such constant -- the dimensions match without one.

The canonical approach to quantum gravity leads to the
Wheeler--DeWitt equation \cite{DeWitt}, which is very like the
time-independent Schr$\ddot{\textrm o}$dinger equation
\cite{CQG94, EOT}. In the case of conformal gravity, it will not
contain any dimensional constant. In the naivest approach, one
will expect a solution of the equation, if it exists, to give
(static) probabilities for different possible shapes of the
configuration of the universe. In the case of the scale-invariant
3-body problem, one should then obtain probabilities for shapes of
triangles. Since the scale-invariant potential for the 3-body
problem has infinitely deep wells at the two-body coincidences and
wave functions tend to collect in potential wells, it could well
be that shapes with one short side of the triangle and two long
ones will get the highest probabilities. These correspond, of
course, to configurations close to a two-particle coincidence. If
so, I would see this as an embryonic theoretical derivation of the
fundamental ratio of scale-invariant quantum cosmology, namely the
ratio of the shortest to the longest length realized in the most
probable configurations. There cannot be a Planck length and a
Hubble radius, but only a shortest-to-longest ratio. It is therefore possible that radical scale invariance as proposed in this paper and in \cite{ABFO} has the potential to derive the Planck length from first principles.

This may also have a bearing on the Hubble red shift. Since it is revealed through quantum transitions, a complete explanation of it must be based on a quantum theory of the universe. This theory is likely to be strongly affected by the key role that the total potential (both gravitational and non-gravitational) plays in the classical theory -- in the `formula for time' (\ref{clem}) it determines how much a given amount of change `advances time' and in (\ref p) it is a factor in the canonical momenta. The potential is therefore likely to have a strong effect on the relative sizes and structures of quantum objects such as atoms and molecules, but these sizes themselves determine the potential. In the classical scale-invariant theory, the nominal initial size of the universe is conserved by the classical equations in the distinguished representation. However, without a quantum form of the theory, we cannot say what we shall measure using physical rods and clocks. Given the central role of the potential in all the basic equations of the theory, it is at least possible that it would affect rods, making them relatively shorter in states of the universe with high absolute magnitude of the potential. This would then mimic an apparent expansion of a universe that becomes more clumpy.

Despite the tentative and speculative nature of these two final sections, I hope they will serve to emphasize the potential
importance of scale invariance. Weyl's Cartesian dream represents unfinished business.

\textbf{Acknowledgements.} I have benefitted greatly from
discussions with the $N$-body specialists Alain Albouy, Alain
Chenciner, J$\ddot{\textrm o}$rg Elsner, Douglas Heggie, Piet Hut, Christian Marchal, Richard
Montgomery, and Carles Sim\'o. Richard suggested 
the use of the moment of inertia to achieve the necessary
homogeneity \cite{RM} and the Alains and Carles showed me what
could be done with the Lagrange--Jacobi relation. Ted Jacobson
also helped in the early stages of this work. I thank John Briggs for discussion of quantum aspects. 
I have also had much
help from my collaborators Edward Anderson, Brendan Foster, and
Niall $\acute{\textrm O}$ Murchadha. The notion of the corrected
coordinate is due in large part to Brendan. I also thank a referee and a member of the editorial board for helpful suggestions for improvement of the first draft.

\section{Appendix: Hamiltonian Formulation}
As Lanczos explains \cite{Lanczos}, p. 169, the Hamiltonian formalism in variational mechanics
is really a special case of the Lagrangian method for which the kinetic energy has a particularly simple 
(canonical) form, namely $T=\sum p^i\dot q_i$. Thus, the (first-order) Lagrangian of the Hamiltonian method is
\begin{equation}
    A=\int\textrm{d}t[\sum p^i\dot q_i-H(q_1,\dots,q_n;p^1,\dots,p^n)],
\label{Lag}
\end{equation}
where $H$ is the normal Hamiltonian and is treated for variational purposes as if it were a potential.

From the Lagrangian (\ref{Lag}), in which the $p^i$'s are coordinates on an equal footing with the 
$q_i$'s except that they appear as Lagrange multipliers (their velocities do not occur in the Lagrangian, 
unlike the $\dot q_i$'s), one calculates the Euler--Lagrange equations in the standard manner. 
They turn out to be the Hamiltonian equations. 

To obtain a Hamiltonian treatment appropriate for scale-invariant gravity, two modifications to the
standard treatment are needed. First, the Lagrange function of any Jacobi action is homogeneous of degree
one in the velocities. This leads to the identity (\ref{QuadCon}), which in turn has the consequence that
when one attempts to calculate the Hamiltonian by the rule $H=p^i\dot q_i-{\cal L}$ an expression that vanishes
identically is obtained. This problem is overcome in Dirac's generalized Hamiltonian theory \cite{Dirac} by converting 
(\ref{QuadCon}) into the quadratic constraint 
$$
{\cal H}=\sum_i{{{\textbf{p}}_{i}}\cdot
    {{\textbf{p}}_{i}}\over 2m_{i}}-(E-U)=0
$$
and taking $H=N\cal H$, where $N$ is a Lagrange multiplier, as the Hamiltonian. In the scale-invariant case,
the constraint is of the same basic form but, first, $E=0$ and, second, $\bar{\cal L}$ is a function of the
corrected coordinates $\bar q_i$ and their velocities. The quadratic constraint (in the one-dimensional case) is
$$
\bar{\cal H}=\sum_i{p^ip^i\over 2m_i}+(a^{\scriptsize\textrm{d}})^2U(\bar q^i)=0,
\hspace{.5cm}\bar q^i=a^{\scriptsize\textrm{d}}(q^i+a^{\scriptsize\textrm{t}}),
$$
where $a^{\scriptsize\textrm{t}}$ and $a^{\scriptsize\textrm{d}}$ are the gauge variables corresponding to translations
and dilatations, and the $(a^{\scriptsize\textrm{d}})^2$ appears in the constraint because 
$p^i=a^{\scriptsize\textrm{d}}{\partial\bar{\cal L}/\partial v^i}$. Any other constraints obtained
in the Lagrangian treatment must be added with further Lagrange multipliers. In our case, we shall need
to add, with appropriate multipliers, the primary constraints (\ref{Niall}) corresponding to the various types 
of best matching. This is relatively standard, through there is, as we shall shortly see, a slight departure from
Dirac's procedure. The second modification concerns the treatment of the gauge auxiliary variables. We need to 
find a Hamiltonian treatment that corresponds to the free-end-point variation in the Lagrangian approach. Both
modifications are achieved if we use the first-order principle
\begin{equation}
    A=\int\textrm{d}\lambda\left(\sum_ip^i\dot q^i
    +p^{\scriptsize\textrm{t}}\dot a^{\scriptsize\textrm{t}}+p^{\scriptsize\textrm{d}}\dot a^{\scriptsize\textrm{d}}-
    N\bar{\cal H}-N^{\scriptsize\textrm{t}}\left(p^{\scriptsize\textrm{t}}-\sum_i p^i\right)-
    N^{\scriptsize\textrm{d}}\left(p^{\scriptsize\textrm{d}}a^{\scriptsize\textrm{d}}-\sum_i p^iq^i\right)\right),
\label{ConstrainedHam}
\end{equation}
where  $p^{\scriptsize\textrm{t}}, p^{\scriptsize\textrm{d}}$ are the canonical momenta corresponding to auxiliary
gauge variables,
and $N^{\scriptsize\textrm{t}},N^{\scriptsize\textrm{d}}$ are multipliers associated with the translational and 
dilatational constraints.

If we now vary (\ref{ConstrainedHam}) with respect to the multipliers $N, N^{\scriptsize\textrm{t}}, 
N^{\scriptsize\textrm{d}}$, we obtain
\begin{equation}
    \bar{\cal H}=\sum_i{p^ip^i\over 2m_i}+(a^{\scriptsize\textrm{d}})^2U(\bar q^i)=0,
     \hspace{.5cm}p^{\scriptsize\textrm{t}}-\sum_i p^i=0,
    \hspace{.5cm}p^{\scriptsize\textrm{d}}a^{\scriptsize\textrm{d}}-\sum_i p^iq^i=0.
\label{PrimaryConstraints}
\end{equation}

We now employ the free-end-point method and require the variation of the action w.r.t both the coordinates
$a^{\scriptsize\textrm{t}}$, $a^{\scriptsize\textrm{d}}$ and the velocities 
$\dot a^{\scriptsize\textrm{t}}$, $\dot a^{\scriptsize\textrm{d}}$ of the auxiliary gauge variables to vanish.
The variation w.r.t the velocities tells us
$$
    p^{\scriptsize\textrm{t}}=0,\hspace{.5cm}p^{\scriptsize\textrm{d}}=0, 
$$
and in conjunction with (\ref{PrimaryConstraints}) these imply
\begin{equation}
    \sum_i p^i=0,\hspace{.5cm}\sum_i p^iq^i=0.
\label{SecCons}
\end{equation}

Variation with respect to the auxiliary variables $a^{\scriptsize\textrm{t}}$ and $a^{\scriptsize\textrm{d}}$ gives
\begin{equation}
    {\partial U\over\partial a^{\scriptsize\textrm{t}}}=-\sum_i{\partial U\over\partial q^i}=0,
    \hspace{.5cm}p^{\scriptsize\textrm{d}}+2a^{\scriptsize\textrm{d}}U+(a^{\scriptsize\textrm{d}})^2
    {\partial U\over\partial a^{\scriptsize\textrm{d}}}=0.
\label{Conds}
\end{equation}

The first of these conditions is the one that ensures Newton's third law, and, since we already have 
$p^{\scriptsize\textrm{d}}=0$, the second gives the homogeneity condition on the potential.  
  
Since the momenta are regarded as independent variables, we also vary w.r.t the momenta of the auxiliary variables,
which gives
\begin{equation}
   \dot a^{\scriptsize\textrm{t}}=N^{\scriptsize\textrm{t}},\hspace{.5cm}
   \dot a^{\scriptsize\textrm{d}}=N^{\scriptsize\textrm{d}}.
\label{Velocities}
\end{equation}

Finally, Hamilton's equations are
\begin{equation}
    \dot q^i={\partial H\over\partial p^i}=N{p^i\over m_i}+N^{\scriptsize\textrm{t}}+N^{\scriptsize\textrm{d}}q^i,
    \hspace{.5cm}\dot p^i=-N(a^{\scriptsize\textrm{d}})^2{\partial U\over\partial q^i}+N^{\scriptsize\textrm{d}}p^i.
\label{HamEqs}
\end{equation}

These reduce to Newton's equations if we go over to the distinguished representation by setting $N=1$ and 
$N^{\scriptsize\textrm{t}}=N^{\scriptsize\textrm{d}}=0$. However, we can perfectly well specify the three multipliers
freely. Then $N^{\scriptsize\textrm{t}}$ and $N^{\scriptsize\textrm{d}}$ are velocities in the unphysical gauge
directions generated by translations and dilatations as in Dirac's treatment \cite{Dirac}, and $N$ is the unphysical `label speed' at which we move along the dynamical orbit in shape space.

However, it will be noted that we have generalized Dirac's treatment by the inclusion of auxiliary gauge variables
in addition to the conventional physical variables and multipliers. The simpler and more familiar form given by Dirac
corresponds to the replacement of (\ref{ConstrainedHam}) by
\begin{equation}
    A^*=\int\textrm{d}\lambda\left(\sum_ip^i\dot q^i-
    N\left(\sum_i{p^ip^i\over 2m_i}+(a^{\scriptsize\textrm{d}})^2U(\bar             q^i)\right)-N^{\scriptsize\textrm{t}}\sum_i p^i-
    N^{\scriptsize\textrm{d}}\sum_i p^iq^i\right).
\label{DiracHam}
\end{equation}
Variation w.r.t the multipliers $N, N^{\scriptsize\textrm{t}},N^{\scriptsize\textrm{d}}$ gives the first member
of (\ref{PrimaryConstraints}) and (\ref{SecCons}). However, the first member of (\ref{Conds}) no longer arises
from the variational principle but must be adjoined if the constraint $\sum_i p^i=0$ is to propagate. In addition,
the status of $a^{\scriptsize\textrm{d}}$ is strange. Variation w.r.t it does give the second member of (\ref{Conds}), but (\ref{Velocities}) has no variational derivation and its second member must be adjoined
in order to propagate the constraint $\bar{\cal H}=0$. We see that in the presence of scale invariance the
Dirac procedure works but lacks a transparent variational basis.

More could be said about these matters, above all about a new unifying perspective on gauge theory that best matching provides. But since this is already a long paper, that must be the subject of another paper.


\begin{thebibliography}{99}
\bibitem{Einstein}  Einstein's objection was reported by Weyl at the end of \cite{Weyl1}
\end{thebibliography}
\end{document}